\documentclass[a4paper,twocolumn,superscriptaddress,10pt,accepted=2019-01-09]{quantumarticle}
\pdfoutput=1
\usepackage[utf8]{inputenc}
\usepackage[english]{babel}
\usepackage[T1]{fontenc}
\usepackage{amsmath}
\usepackage{amsfonts}
\usepackage{amsthm}
\usepackage{amssymb}
\usepackage{hyperref}
\hypersetup{
colorlinks=true,
linkcolor=red,
citecolor=red}

\usepackage[sort&compress,numbers]{natbib}

\usepackage{textcomp}
\usepackage{graphicx}
\usepackage{enumerate}
\usepackage{multirow}
\usepackage{tabularx}
\usepackage{mathtools}
\usepackage[mathscr]{euscript}
\usepackage{dsfont}
\usepackage[normalem]{ulem}
\usepackage{centernot}
\usepackage{longtable}

\usepackage{stmaryrd}
\SetSymbolFont{stmry}{bold}{U}{stmry}{m}{n}

\newtheorem{lemma}{Lemma}

\newcommand\dmu[1]{\int\frac{d^{2} #1}{\pi}}
\newcommand\ddmu[2]{\int\frac{d^{2}#1 d^2#2}{\pi^2}}
\newcommand\s[1]{_{\rm #1}}

\newcommand{\bra}[1] {\langle #1 |}
\newcommand{\ket}[1] {| #1 \rangle}

\newcommand{\one}{\leavevmode\hbox{\small1\normalsize\kern-.33em1}}

\begin{document}

\title{Optomechanical state reconstruction and nonclassicality \quad verification beyond the resolved-sideband regime}

\date{\today}
\author{Farid Shahandeh}
\orcid{0000-0001-5316-6707}
\email{shahandeh.f@gmail.com}
\affiliation{Centre for Quantum Computation and Communication Technology, School of Mathematics and Physics, University of Queensland, St Lucia, Queensland 4072, Australia}
\affiliation{Department of Physics, Swansea University, Singleton Park, Swansea SA2 8PP, United Kingdom}
\author{Martin Ringbauer}
\orcid{0000-0001-5055-6240}
\affiliation{Centre for Engineered Quantum Systems, School of Mathematics and Physics, University of Queensland, St Lucia, Queensland 4072, Australia}
\affiliation{Institute for Experimental Physics, University of Innsbruck, Technikerstra\ss e 25, 6020 Innsbruck, Austria}

\maketitle

\begin{abstract}
Quantum optomechanics uses optical means to generate and manipulate quantum states of motion of mechanical resonators. This provides an intriguing platform for the study of fundamental physics and the development of novel quantum devices. Yet, the challenge of reconstructing and verifying the quantum state of mechanical systems has remained a major roadblock in the field. Here, we present a novel approach that allows for tomographic reconstruction of the quantum state of a mechanical system without the need for extremely high quality optical cavities. We show that, without relying on the usual state transfer presumption between light an mechanics, the full optomechanical Hamiltonian can be exploited to imprint mechanical tomograms on a strong optical coherent pulse, which can then be read out using well-established techniques. Furthermore, with only a small number of measurements, our method can be used to witness nonclassical features of mechanical systems without requiring full tomography. By relaxing the experimental requirements, our technique thus opens a feasible route towards verifying the quantum state of mechanical resonators and their nonclassical behaviour in a wide range of optomechanical systems.
\end{abstract}

\maketitle

\section{Introduction}
Optomechanics~\citep{Aspelmeyer2014} where a mechanical oscillator interacts with an optical field via radiation pressure, is a promising direction of research for fundamental physics~\citep{Bose1999, Marshall2003, Pikovski2012} and the development of novel weak-force sensors~\citep{Rugar2004}. Yet, preparing a large mechanical device in a quantum state of motion, and then verifying and exploiting the quantum nature of such a system have remained elusive goals. Most research to date has focused on the problem of state preparation and there have been significant recent advances using measurement-based techniques to overcome the limitations of weak coupling and non-zero initial thermal occupation~\citep{Vanner2013,Ringbauer2016Optomechanics}. These techniques bring the preparation of non-classical states of motion within the realm of current experimental capabilities. Here, we thus focus on the problem of observing quantum states of motion.

Continuous-variable quantum systems, such as optical fields or mechanical resonators, are best described using distribution functions over a quantum phase-space spanned by two quadratures of interest, such as position and momentum, or amplitude and phase. However, one of the distinctive features of quantum systems is that any such distribution must allow for negative values and can thus not be a bona fide probability measure. The class of viable distributions form a single-parameter family, called the \emph{$s$-parameterized quasiprobability distributions}~\citep{GlauberBook}. Of particular interest within this family is the \emph{Wigner function}, corresponding to a value of $s=0$, as the only distribution that faithfully represents the quantum state and, at the same time, correctly reproduces the marginal quadrature distributions. Another important case is the \emph{P-function}, corresponding to $s=1$, whose negativity is one of the main signatures of nonclassical behaviour~\citep{VogelBook}. 

In principle, the Wigner function of an unknown quantum state $\hat{\varrho}$ can be reconstructed from quadrature measurements via an inverse Radon transform. This technique has been demonstrated for optical fields~\citep{Lvovsky2001,Zavatta2004}, where quadrature measurements are readily available using homodyne detection~\citep{Leonhardt1997,Hansen2001}. For mechanical resonators, however, direct quadrature measurements are not available. Instead, the common approach aims to use a high-quality optical cavity in the \emph{resolved sideband} regime, which would allow to transfer the mechanical state onto an optical field, which can be reconstructed using established techniques. This regime, however, is experimentally very difficult to reach~\citep{Vanner2015}. Techniques that work outside the demanding resolved sideband regime, where state transfer in the usual sense is not available, are thus highly desirable. One such scheme, using a classical optical readout field, has been proposed and experimentally implemented recently~\citep{Vanner2011,Vanner2013}. Under realistic conditions, however, limitations from optical shot noise of the readout field ultimately preclude an experimental reconstruction of the mechanical Wigner function using this approach~\citep{Vanner2015}. Consequently, Wigner-function tomography of a mechanical system remains an outstanding challenge.

Here, we introduce an alternative approach which works outside the resolved sideband regime with no, or only a low quality cavity. Our technique works by imprinting the mechanical quadrature distributions on a strong optical pulse in a coherent or squeezed state, from which it can be extracted using standard optical methods. We show that this makes it possible to overcome the noise limitations of previous approaches and perform a tomographic reconstruction of any $s$-parameterized quasiprobability distribution, including the Wigner function of the mechanical motion. Furthermore, since our technique is independent of the single-photon coupling strength it is suitable for a wide range of optomechanical systems with current technology. 
Finally, we discuss how our approach could be used to detect P-function nonclassicality of a mechanical system with only a small number of measurements. We also prove the robustness of our nonclassicality criteria against readout noise and detection inefficiency.

\section{Results} 
This work is organised as follows. 
We first provide an overview of the optomechanical interaction without assuming resolved mechanical sidebands.
We then discuss the description of this interaction over the quantum mechanical phase-space. There we also derive the required transformations, and present our readout scheme for obtaining the mechanical phase-space distributions as the main result of this paper. We then compare our scheme with the existing techniques in the non-resolved sideband regime. Finally, we provide a necessary condition for demonstrating mechanical nonclassicality using the smallest subset of data obtained from our scheme, namely, a single mechanical tomogram. We further show that this condition is robust against noise and detection inefficiency, and describe an explicit experimental protocol which we show can be implemented with current technology in a variety of optomechanical systems.

\subsection{Optomechanical interaction}
We now consider a mechanical resonator coupled to an optical field via the radiation pressure interaction. A single photon reflecting off a mechanical resonator imparts a momentum kick proportional to the photon frequency to the resonator, while obtaining a phase-shift proportional to the mechanical position. The evolution of the joint system is then described by the Hamiltonian
\begin{equation}
\label{eq:Hamilton}
\frac{1}{\hbar}\hat{H}= \omega\s{o}a^\dag a + \omega\s{m}b^\dag b + g_0 a^\dag a(b^\dag+b) ,
\end{equation}
where $a$ ($b$) is the optical (mechanical) bosonic annihilation operator. The terms of the Hamiltonian correspond, respectively, to the free evolution of the optical field with frequency $\omega\s{o}$, the free evolution of the mechanics with frequency $\omega\s{m}$, and the radiation pressure interaction. The latter couples the mechanical position operator $\hat X\propto b^\dag + b$ to the optical photon number operator $\hat N = a^\dag a$ with the single photon coupling strength $g_0=Gx\s{zpf}$, where $G$ is the coupling constant, and $x\s{zpf}=\sqrt{\hbar/2m\s{eff}\omega\s{m}}$ the mechanical zero-point fluctuation. In this picture the mechanical response to a single photon is captured by $g_0$, and can be linearly enhanced by a cavity that allows for multiple interactions per photon~\citep{Aspelmeyer2014}.

In the cavity-enhanced case one can identify a number of interesting parameter regimes. Much work to date has focused on the so-called \emph{resolved sideband regime}, where the decay rate of the optical cavity $\kappa_\text{o}$ is much smaller than the mechanical frequency. This makes it possible to excite, to good approximation, only one of the two motional sidebands of the cavity resonance. Ignoring non-resonant terms, which is known as the rotating wave approximation, the optomechanical interaction is then dominated, either by a beam-splitter-like interaction for the red sideband, or a squeezing-like interaction for the blue sideband~\citep{AspelmeyerBook}.
For tomography, however, this regime is challenging due to the incompatible requirements of a high optical quality factor for sideband resolution and a low optical quality factor to retain control over the optical amplitude fluctuations inside the cavity. 

\begin{figure*}[t]
  \includegraphics[width=\textwidth]{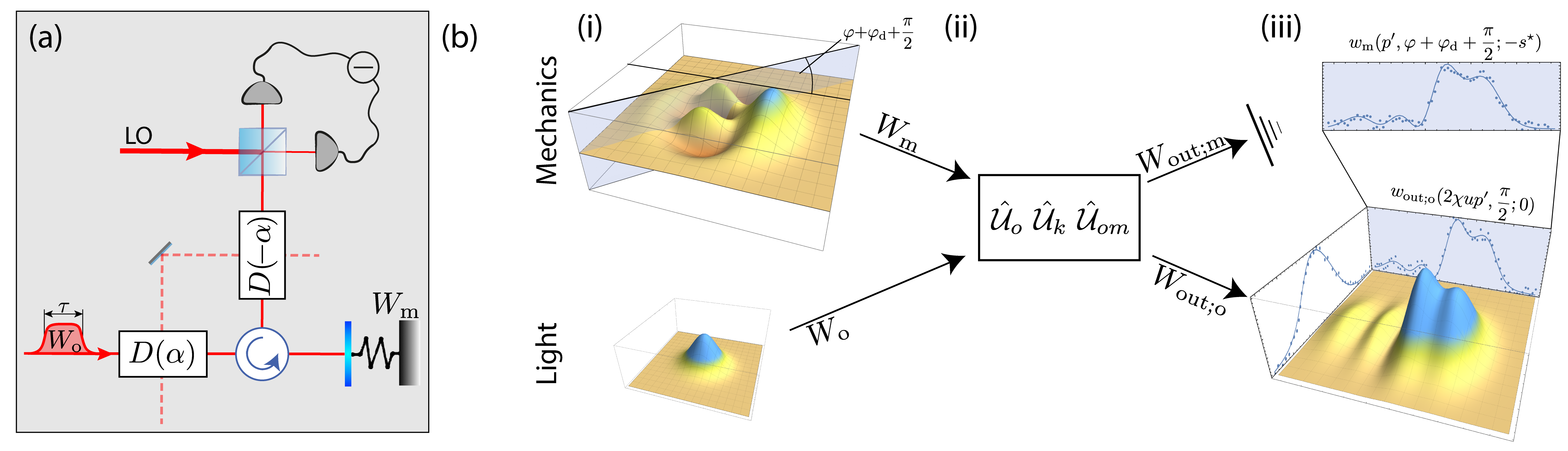}
  \vspace{-1em}
  \caption{\textbf{Sketch of our tomography method.} \textbf{(a)}~Schematic experimental setup. The initial (squeezed) vacuum state $\mathscr{W}\s{o}$ is first displaced by $\alpha$ via interference with a strong laser pulse of length $\tau$ (dashed line) on a highly transmissive beamsplitter. The probe beam then interacts with the mechanical system, is subsequently displaced by $-\alpha$, and finally measured using homodyne detection. Both displacements are easily implemented in an inherently stable manner using the same laser pulse, see Secs.~\ref{sec:imperfections} and \ref{sec:experiment} for details.
  \textbf{(b)} In the phase-space representation (i)~the mechanical resonator is initially in an unknown state $\mathscr{W}\s{m}$ and the optical system is in the state $\mathscr{W}\s{o}$. (ii)~The two systems interact under the full optomechanical Hamiltonian, in the case of a strong optical displacement given by Eqs.~\eqref{eq:OMint}--\eqref{eq:Oint}. (iii)~After the interaction, the $s$-parameterized $\phi$-tomogram of the mechanical system will be imprinted on the momentum-quadrature of the optical Wigner-function, which can be obtained from a standard homodyne measurement. Using a number of such optical/mechanical tomograms one can use the inverse Radon transformation in Fig.~\ref{fig:Motivation} to reconstruct the mechanical $s$-parameterized quasiprobability distribution. For illustrative purposes only we chose an initial mechanical state of the form $\left(\ket{0}+2\ket{1}+\ket{2}+2\ket{3}\right)/\sqrt{10}$ in the Fock-state basis. The displayed tomograms include points where we numerically added $10\%$ noise to indicate measurement imprecision.}
    \label{fig:Tomography}
\end{figure*}

Here, we are interested in the technically simpler scenario in which the optical field is a free-space mode, or weakly enhanced by a low-quanlity factor optical cavity (with a cavity decay rate much faster than the mechanical frequncy), see Fig.~\ref{fig:Tomography}a. In this regime, the motional sidebands cannot be resolved, implying that the rotating-wave approximation is not valid and the full optomechanical Hamiltonian in Eq.~\eqref{eq:Hamilton} must be considered.
Suppose that the mechanical system is initially in a state $\hat{\varrho}\s{m}$ which then interacts with a displaced optical pulse, $\hat{D}(\alpha)\hat{\varrho}\s{o}\hat{D}^\dag(\alpha)$ over a duration $\tau$. At a time $\tau_0$ after the interaction, the optical field is displaced by $-\alpha$. The evolution of the state is thus given by $\hat{D}(-\alpha)\exp{-i \tau_0 (a^\dag a + b^\dag b)}\exp\{-i \tau [\omega\s{o}a^\dag a + \omega\s{m}b^\dag b + g_0 a^\dag a(b^\dag+b)]\} \hat{D}(\alpha)$.
As we show in detail in Appendix~\ref{app:Disentangle}, under the condition that the optical field is strong (i.e.\ $|\alpha|\gg 1$) and using the fact that we only perform measurements on the optical field, the evolution of the system can be disentangled and linearised to $\hat{\mathcal{U}}\s{o}~\hat{\mathcal{U}}\s{k}~\hat{\mathcal{U}}\s{om}$, where $\hat{\mathcal{U}}\s{om}$, $\hat{\mathcal{U}}\s{k}$, and $\hat{\mathcal{U}}\s{o}$ are unitaries representing the optomechanical interaction, an optical Kerr interaction, and a displaced optical free evolution, respectively, given by
\begin{align}
& \hat{\mathcal{U}}\s{om} = e^{\chi u(\tilde{a}^\dag_\theta + \tilde{a}_\theta)(\tilde{b}^\dag_\varphi - \tilde{b}_\varphi)}, \label{eq:OMint}\\
& \hat{\mathcal{U}}\s{k} = e^{iv[4 a^\dag a + 2r(e^{i\theta}a^\dag + e^{-i\theta}a) + (e^{2i\theta} a^{\dag 2} + e^{-2i\theta}a^2)]}, \label{eq:Kint}\\
& \hat{\mathcal{U}}\s{o} = \hat{D}(-\alpha) e^{-i (\tau_0 + \tau)  \omega\s{o}a^\dag a} \hat{D}(\alpha).\label{eq:Oint}
\end{align}
Here, we have defined the parameters as follows:
$\alpha = re^{i\theta}$ with $r\in\mathbb{R}^+$, $\mu = 1-e^{i\omega\s{m}\tau} = ue^{i\varphi}$ with 
\begin{equation}\label{eq:mecphase}
u{=}\sqrt{2(1{-}\cos{\omega\s{m}\tau})},\quad
\varphi{=}\tan^{-1}{\Big(\frac{\sin{\omega\s{m}\tau}}{1-\cos{\omega\s{m}\tau}}\Big)},
\end{equation}
and
\begin{align}
& \chi = \frac{g_0 r}{\omega\s{m}},\label{eq:chi}\\
& \tilde{a}_\theta =ae^{-i\theta}+\frac{r}{2}, \quad \tilde{b}_\varphi = b e^{-i\varphi}.
\end{align}
It is important to note that, while the phase of the optical mode $\theta$ is being set by the input optical coherent state, the relevant mechanical phase $\varphi$ is determined by the length of the input optical pulse $\tau$.

\subsection{Optomechanical interaction in phase space} \label{sec:OMintPS}
\subsubsection{Mode transformation}

By defining a vector of input mode operators $\hat{\vec{A}}:=(a^\dag, b^\dag, a, b)$, we aim to find the the output mode operators in the Heisenberg picture under the aforementioned unitary evolution, i.e.\ $\hat{\vec{A}}'=(\hat{\mathcal{U}}\s{o}\hat{\mathcal{U}}\s{k}\hat{\mathcal{U}}\s{om})\hat{\vec{A}}(\hat{\mathcal{U}}^\dag\s{om}\hat{\mathcal{U}}^\dag\s{k}\hat{\mathcal{U}}^\dag\s{o})$.
Since these unitary evolutions are single-mode and two-mode linear transformations, they correspond to elements of the symplectic groups ${\rm Sp}(2,\mathbb{R})$ and ${\rm Sp}(4,\mathbb{R})$, respectively. We can thus exploit the properties of symplectic transformations, in particular, that if $\hat{\mathcal{U}}=\exp\{\frac{1}{2}\hat{\vec{A}} \ln (M) \Sigma \hat{\vec{A}}^\mathsf{T} \}$, with $\Sigma={\rm antidiag}\{I,-I\}$, then $\hat{\mathcal{U}}\hat{\vec{A}}\hat{\mathcal{U}}^\dag= \hat{\vec{A}} M$~\citep{Wang1994,Zhang1994}.
Using Eqs.~\eqref{eq:OMint},~\eqref{eq:Kint}, and~\eqref{eq:Oint}, after some algebra and making reasonable restrictions on the choice of the optical pulse $\tau$ and the delay duration $\tau_0$, via 
\begin{equation} \label{eq:cond}
\begin{matrix*}[l]
\omega\s{m}\tau  - \sin(\omega\s{m}\tau ) = k\pi/2\sqrt{3} \chi^2, & \quad k \in 2\mathbb{Z}^+,\\
(\tau_0 + \tau)\omega\s{o} = 2\pi m, & \quad m \in \mathbb{N},
\end{matrix*}
\end{equation}
the transformed optomechanical mode operators are found as 
\begin{equation}
\label{eq:aprime}
\begin{split}
& a' = a - i\sqrt{2}\chi u \hat{X}(\varphi + \frac{\pi}{2}),\\
& b' = b - \sqrt{2} \chi u e^{i\varphi} [\hat{x}+r].
\end{split}
\end{equation}
Here, $\hat{X}(\varphi)=(b^\dag e^{i\varphi} + b e^{-i\varphi})/\sqrt{2}$ and $\hat{x} = (a^\dag + a )/\sqrt{2}$ are the $\varphi$-quadrature of the mechanics and the $x$-quadrature of the optical field, respectively. We refer the interested reader to Appendix~\ref{app:Transes} for the detailed derivation of the general mode transformation, as well as the simplified Eqs.~\eqref{eq:cond} and~\eqref{eq:aprime}. Importantly, in Eq.~\eqref{eq:cond} we require that $0\ll\omega\s{m}\tau\ll 2\pi$, both to obtain a nonzero coupling ($u\neq 0$) and to minimize the effect of mechanical losses. This can be achieved by assuming the pulse duration $\tau$ to be on the order of half a mechanical period, i.e., $\omega\s{m}\tau \approx \pi$. It is evident from Eq.~\eqref{eq:aprime} that the output optical field now carries information about the mechanical quadratures. At the same time, the mechanical mode is affected by the back-action noise proportional to the input light quadrature. Before we can proceed to the Wigner-function picture, we need the following result, the proof of which is given in the Appendix~\ref{app:ProofLem}. 
\begin{lemma}\label{lemma1}
For the order parameter $s=0$, a product of Weyl-Wigner operators remains a product under all linear transformations of the mode operators, $
\begin{pmatrix}
a'^\dag & b'^\dag & a' & b'
\end{pmatrix}
=
[
\begin{pmatrix}
a^\dag & b^\dag & a & b
\end{pmatrix}
S
+
\begin{pmatrix}
D\s{a}^* & D\s{b}^* & D\s{a} & D\s{a}
\end{pmatrix}
]$,
the argument of which changes according to the inverse of the corresponding symplectic transformation, i.e., $\hat{T}(\alpha;0)\otimes\hat{T}(\beta;0)=\hat{T}(\alpha';0)\otimes\hat{T}(\beta';0)$ where $\begin{pmatrix}
\alpha'^* & \beta'^* & \alpha' & \beta'
\end{pmatrix}
=
[
\begin{pmatrix}
\alpha^* & \beta^* & \alpha & \beta
\end{pmatrix}
-
\begin{pmatrix}
D\s{a}^* & D\s{b}^* & D\s{a} & D\s{a}
\end{pmatrix}
]
S^{-1}$.
\end{lemma}

\subsubsection{Interaction in phase space }
We are now ready to examine the optomechanical interactions in the phase-space representation, see Fig.~\ref{fig:Tomography}b.
The Wigner decomposition (see Sec.~\ref{sec:PSRep}) of the joint initial optomechanical state is
\begin{equation*}
\hat{\varrho}\s{in}=\ddmu\alpha\beta \mathscr{W}\s{o}(\alpha;0)\mathscr{W}\s{m}(\beta;0) \hat{T}\s{o}(\alpha;0)\otimes\hat{T}\s{m}(\beta;0) ,
\end{equation*}
where $\mathscr{W}\s{o}$ and $\mathscr{W}\s{m}$ are the Wigner functions of light and mechanics, respectively.
Allowing for a free mechanical evolution for a time $\tau\s{d}$ before the interaction with the optical field, the output state is given by $\hat{\varrho}\s{out}=[\hat{\mathcal{U}}\s{o}\hat{\mathcal{U}}\s{k}\hat{\mathcal{U}}\s{om}\hat{\mathcal{R}}\s{m}(\varphi\s{d})]\hat{\varrho}\s{in}[\hat{\mathcal{R}}^\dag\s{m}(\varphi\s{d})\hat{\mathcal{U}}^\dag\s{om}\hat{\mathcal{U}}^\dag\s{k}\hat{\mathcal{U}}^\dag\s{o}]$ where $\hat{\mathcal{R}}\s{m}(\varphi\s{d})=\exp\{-i\varphi\s{d}b^\dag b\}$ with $\varphi\s{d}=\omega\s{m}\tau\s{d}$. 
This results in $\hat{\varrho}\s{out} = \ddmu\alpha\beta \mathscr{W}\s{o}(\alpha;0)\mathscr{W}\s{m}(\beta;0) \hat{T}\s{o}(\alpha';0)\otimes\hat{T}\s{m}(\beta';0)$.
Using Lemma~\ref{lemma1} and the inverse of the symplectic transformation corresponding to the overall evolution $\hat{\mathcal{U}}\s{o}\hat{\mathcal{U}}\s{k}\hat{\mathcal{U}}\s{om}$ given by Eq.~\eqref{eq:aprime}, we obtain 
\begin{equation*}
\begin{split}
& \alpha' = \alpha + i\sqrt{2}\chi u X(\varphi+\varphi\s{d}+\frac{\pi}{2}),\\
& \beta' = \beta e^{-i\varphi\s{d}} + \sqrt{2}\chi u e^{i\varphi} [x(\theta)+r].
\end{split}
\end{equation*}
Since we only perform measurements on the optical field, we can trace out the mechanics and make use of the normality of the Weyl-Wigner operators to get $\hat{\varrho}\s{out;o} = \ddmu\alpha\beta \mathscr{W}\s{o}(\alpha;0)\mathscr{W}\s{m}(\beta;0) \hat{T}\s{o}(\alpha';0)$.
Using the orthogonality relation of the Weyl-Wigner operators, the final measured Wigner-function of the optical field can be readily obtained as $\mathscr{W}\s{out;o}(\alpha';0) =\dmu\beta \mathscr{W}\s{o}[\alpha' - i\sqrt{2}\chi u X(\varphi+\varphi\s{d}+\pi/2);0]\mathscr{W}\s{m}(\beta;0)$. Switching to position and momentum coordinates $(x',p')$ of the optical output, we obtain
\begin{equation}\label{eq:tomomain1}
\begin{split}
&\mathscr{W}\s{out;o}(x',p';0) =\\
&\dmu\beta \mathscr{W}\s{o}[x' , p' - 2\chi u X(\varphi+\varphi\s{d}+\frac{\pi}{2});0]\mathscr{W}\s{m}(\beta;0),
\end{split}
\end{equation}
showing that only the momentum quadrature of the optical output is modified by the interaction.

\subsection{Mechanical tomography}
Having established the tools to investigate the information about the mechanical quantum state imprinted in the refelcted probe ligh, we now describe how to use these tools to achieve quantum tomography of the mechanical state as depicted in Fig.~\ref{fig:Tomography}b.

\subsubsection{General case readout} \label{sec:Gread}
In principle, an optical vacuum state would be sufficient as a probe pulse for our method. However, in the general case we now consider a momentum-squeezed vacuum state with Wigner function 
\begin{equation}\label{eq:SVS}
\mathscr{W}\s{o}(x,p;0) = \frac{1}{\pi}\exp\{-(e^{-2\varepsilon}x^2 + e^{2\varepsilon} p^2)\},
\end{equation}
where $\varepsilon\in[0,\infty)$ is the squeezing parameter. While this is not strictly necessary for our protocol, it adds an additional parameter that could be convenient for some experimental implementations. 
Using this initial state in Eq.~\eqref{eq:tomomain1} and changing the variable $p'\rightarrow 2\chi u p'$, we obtain the Wigner function of the optical output as
\begin{equation}\label{eq:GkernelTomo}
\begin{split}
&\mathscr{W}\s{out;o}(x',2\chi u p';0) = \frac{1}{\pi} e^{-e^{-2\varepsilon}x'^2}\\
&\quad \times \dmu\beta e^{-4\chi^2 u^2 e^{2\varepsilon} [p' - X(\varphi+\varphi\s{d}+\frac{\pi}{2})]^2} \mathscr{W}\s{m}(\beta;0).
\end{split}
\end{equation}
From Eq.~\eqref{eq:GkernelTomo} it is evident that the information about the mechanical $(\varphi+\varphi\s{d}+\pi/2)$-quadrature is encoded within the momentum quadrature of the optical field. 
Hence, we can integrate over the position of the optical Wigner function and arrive at
\begin{equation}\label{eq:OtomoMtomo1}
\begin{split}
&w\s{out;o}(2\chi u p',\frac{\pi}{2}) =  \\
& \qquad \frac{e^{\varepsilon}}{\sqrt{\pi}} \dmu\beta e^{-4\chi^2 u^2 e^{2\varepsilon} [p' - X(\varphi+\varphi\s{d}+\frac{\pi}{2})]^2} \mathscr{W}\s{m}(\beta;0),
\end{split}
\end{equation}
in which $w\s{out;o}(p,\frac{\pi}{2})$ is the momentum tomogram of the optical output pulse.
Interestingly, we can identify the Gaussian kernel in Eq.~\eqref{eq:OtomoMtomo1} with the $s$-parameterized quasiprobability of the tomographic operator (see Sec~\ref{sec:PSTomo}), obtaining
\begin{equation}
\begin{split}
& e^{-4\chi^2 u^2 e^{2\varepsilon} [p' - X(\varphi+\varphi\s{d}+\frac{\pi}{2})]^2} = \\
& \qquad \qquad \qquad \sqrt{s^\star\pi}~\mathscr{W}_{\hat{\Pi}(p',\varphi+\varphi\s{d}+\frac{\pi}{2})}(\beta;-s^\star),
\end{split}
\end{equation}
with order parameter
\begin{equation}
\label{eq:IntOrder}
s^\star = (2\chi u e^{\varepsilon})^{-2}.
\end{equation}
As we show in Appendix~\ref{app:OrderShift}, the order mismatch can be moved to the mechanical state, which leads to
\begin{equation}
\begin{split}
& \dmu\beta e^{-4\chi^2 u^2 e^{2\varepsilon} [p' - X(\varphi+\varphi\s{d}+\frac{\pi}{2})]^2} \mathscr{W}\s{m}(\beta;0) =\\
& \sqrt{s^\star\pi}\dmu\beta \mathscr{W}_{\hat{\Pi}(p',\varphi+\varphi\s{d}+\frac{\pi}{2})}(\beta;-s^\star) \mathscr{W}\s{m}(\beta;0) = \\
& \sqrt{s^\star\pi}\dmu\beta \mathscr{W}_{\hat{\Pi}(p',\varphi+\varphi\s{d}+\frac{\pi}{2})}(\beta;0) \mathscr{W}\s{m}(\beta;-s^\star) =\\
& \qquad \sqrt{s^\star\pi} ~w\s{m}(p',\varphi+\varphi\s{d}+\frac{\pi}{2};-s^\star) .
\end{split}
\label{eq:tomokershift}
\end{equation}
Here, we have defined the $s$-parameterized mechanical tomogram $w\s{m}(x,\varphi;s)$ to be the marginal of the $s$-parameterized quasiprobability distribution of the mechanical quantum state. 
Importantly, the relation obtained in~\eqref{eq:tomokershift} represents a Radon transformation (see Sec.~\ref{sec:PSTomo}) of the $(-s^\star)$-parameterized quasiprobability distribution $\mathscr{W}\s{m}(\beta;-s^\star)$.
\begin{equation}
\begin{split}
&\mathscr{W}\s{m}(\beta;-s^\star) \quad \xleftrightarrow[\text{Inv. Radon transform}]{\text{Radon transform}}\quad \{w\s{m}(x,\phi;-s^\star)\},
\end{split}
\end{equation}
Substituting Eq.~\eqref{eq:tomokershift} into Eq.~\eqref{eq:OtomoMtomo1} we find
\begin{equation}\label{eq:GenTomo}
\begin{split}
w\s{out;o}(2\chi u p',\frac{\pi}{2};0) = \frac{1}{2\chi u} w\s{m}(p',\varphi+\varphi\s{d}+\frac{\pi}{2};-s^\star),
\end{split}
\end{equation}
where $s^\star$ is given by Eq.~\eqref{eq:IntOrder}.

Equation~\eqref{eq:GenTomo} is the main result of this manuscript, implying that the $(-s^\star)$-parameterized mechanical tomogram is encoded, up to a scaling factor in the momentum quadrature of the optical Wigner function, which can be easily read out using well-established techniques. Note, however, that due to the scaling one might require much larger quadrature detection-range in the optical homodyne measurement to access the optical tomograms over the required range of values. This can be avoided by setting $2\chi u =1$ and instead exploiting the momentum squeezing of the initial optical state to set the order parameter $s^\star$. 
We will discuss in detail the experimental feasibility of all such assumptions in our experimental proposal.

\subsubsection{Strong interaction regime}
 Recalling Eq.~\eqref{eq:IntOrder}, it turns out that the order parameter $s^\star$ goes to zero quadratically with the displacement amplitude $r$ and exponentially with the squeezing parameter $\varepsilon$. Hence, for sufficiently large $\chi u e^{\varepsilon}$ we can approximate
\begin{equation}\label{eq:orderApprox}
\begin{split}
w\s{m}(p',\varphi+\varphi\s{d}+\frac{\pi}{2};-s^\star) &\approx w\s{m}(p',\varphi+\varphi\s{d}+\frac{\pi}{2};0)
\end{split}
\end{equation}
which in turn, implies that the $\phi$-tomogram of the mechanical \emph{Wigner function} is imprinted in a rescaled form on the momentum quadrature of the optical output Wigner function,
\begin{equation}
\label{eq:Mtomogram}
\begin{split}
w\s{out;o}(2\chi u p',\frac{\pi}{2};0) = \frac{1}{2\chi u}  w\s{m}(p',\varphi+\varphi\s{d}+\frac{\pi}{2};0).
\end{split}
\end{equation}

\subsubsection{Advantages over standard pulsed tomography}
The use of short pulses of coherent light for optomechanical state tomography outside the resolved sideband regime was first proposed in Ref.~\citep{Vanner2011}. Their approach works by imprinting the mechanical quadrature distribution onto the momentum quadrature of the probe pulse~\citep{Vanner2015}, which has recently been demonstrated experimentally~\citep{Vanner2013}. In practice, however, this simple technique is limited by the fundamental quantum noise on the readout pulse. This precludes experimental access to the Wigner function of the mechanical system and indeed might not allow for the reconstruction of any quasiprobability distribution with $s>-1$, as we discuss below.

The approach of Ref.~\citep{Vanner2011,Vanner2013} uses classical readout pulses, which have a non-negative quasiprobability distribution for $s=1$. The bosonic coherent state basis in this case is given by $\hat{T}(\alpha;-1)=\ket{\alpha}\bra{\alpha}$, and thus
\begin{equation}\label{eq:Pfunc}
\hat{\varrho} = \dmu\alpha P(\alpha) \ket{\alpha}\bra{\alpha} .
\end{equation}
The distribution corresponding to the optical field is then a Gaussian of the form
\begin{equation}\label{eq:classicalP}
P(\alpha)=\frac{1}{2\pi\sigma_x\sigma_p}\ e^{-\frac{(x-\bar{x})^2}{2\sigma^2_x}-\frac{(p-\bar{p})^2}{2\sigma^2_p}} ,
\end{equation}
representing a (possibly squeezed) thermal state of light with $x=\sqrt{2}\,{\rm Re}\alpha$ and $p=\sqrt{2}\,{\rm Im}\alpha$. 
The function $P(\alpha)$ is non-negative and corresponds to a class of quantum states called P-classical states.
The method of Ref.~\citep{Vanner2015} can now be recast in our notation as
\begin{equation}\label{eq:MVRel1}
\begin{split}
w\s{out;o}(p,\frac{\pi}{2};0) = \int\frac{dx}{\sqrt{\pi}} e^{-\frac{(p-\eta x)^2}{1+2\sigma_p^2}} w\s{m}(x,\varphi;0)
\end{split}
\end{equation}
or equivalently,
\begin{equation}\label{eq:MVRel2}
w\s{out;o}(\eta p,\frac{\pi}{2};0) = \int\frac{dx}{\sqrt{\pi}} e^{-\frac{\eta^2(p-x)^2}{1+2\sigma_p^2}} w\s{m}(x,\varphi;0) ,
\end{equation}
where $\eta$ is the strength of the mechanical position measurement.
Despite similarities between Eq.~\eqref{eq:MVRel2} and our Eq.~\eqref{eq:OtomoMtomo1}, there are important differences providing an advantage to our method over that of Ref.~\citep{Vanner2011,Vanner2015}.
First, by assuming the input to be P-classical, one imposes the restriction that the variance of the momentum quadrature of the readout is shot-noise limited, i.e.\ $\sigma_p$ in Eq.~\eqref{eq:classicalP} and hence in~Eq.~\eqref{eq:MVRel2} is always non-negative.
Thus, the sharpest mechanical quadrature measurement possible within that scheme is achieved in the limit $\sigma_p\to0$, corresponding to an optical coherent state readout.
Consequently, nonclassical states, in particular those with negative P-functions that possess ill-defined P-function variances, e.g.\ squeezed coherent states, are not considered in this approach.
In contrast, our method makes no assumption about the classicality of the readout optical field to arrive at Eqs.~\eqref{eq:OtomoMtomo1} and~\eqref{eq:GenTomo}, allowing us to use a nonclassical readout such as a squeezed coherent state of light. This makes it possible to use an optical readout field with a momentum variance significantly below the shot-noise limit.

To extract the mechanical Wigner tomograms $w\s{m;o}(x,\varphi;0)$ from the output optical tomograms $w\s{out;o}(p,\frac{\pi}{2};0)$ using Ref.~\citep{Vanner2011,Vanner2015}, the Gaussian convolution kernel $\exp\{-\eta^2(p-x)^2/(1+2\sigma_p^2)\}$ on the r.h.s.\ of Eq.~\eqref{eq:MVRel2} must be as narrow as possible to be approximated by a Dirac delta function. Using the best classical readout with $\sigma_p=0$, there are two possibilities to achieve this goal.
\begin{enumerate}[(i)]
\item Increasing the strength of the readout pulses such that $\eta\gg 1$~\citep{Vanner2013}. This route, however, is problematic, since the scaling on the l.h.s.\ of Eq.~\eqref{eq:MVRel2} implies that increasing $\eta$ also requires larger sensitivity in the optical homodyne measurement to access the optical tomograms over the required range of values. Hence, it is desirable to work in a regime where $\eta \approx 1$, leaving us with the second possibility below.
\item Inverting the convolution on the r.h.s.\ of Eq.~\eqref{eq:MVRel2} by applying a deconvolution map to the output optical tomogram $w\s{out;o}(\eta p,\frac{\pi}{2};0)$ with the kernel $e^{\eta^2(p-x)^2}$. In theory this allows one to compute the mechanical tomogram $w\s{m}(x,\varphi;0)$ exactly. In practice, however, the numerical process becomes very unstable when applied to real experimental data~\citep{Vanner2015}, since the exponentially rising kernel strongly amplifies any noise~\citep{Wallentowitz1996}.
\end{enumerate}

In contrast, our approach opens up a third possibility to overcome the smoothing problem above, namely by using a \emph{nonclassical} readout pulse. A careful comparison between Eq.~\eqref{eq:MVRel2} and our Eq.~\eqref{eq:OtomoMtomo1} shows that the extra free parameter $\varepsilon$ introduced into our model via squeezing of the initial optical probe neither acts as an scaling factor nor puts any strict lower bounds on the variance of the momentum quadrature of the readout. Consequently, using our approach one can reliably set the scaling parameter $2\chi u \approx 1$, and manipulate the momentum squeezing to achieve, in principle, an ordering parameter $s^\star$ arbitrarily close to zero. Our approach thus allows for an enhanced mechanical quadrature measurement without the previous difficulties, thereby overcoming one of the most challenging problems in optomechanical tomography through the use of momentum-squeezed probe light.

\subsection{Certification of mechanical nonclassicality} \label{sec:NCverify}
The method introduced above allows for the full reconstruction of $s$-parametrized quasiprobability distributions of a mechanical system in a new experimental regime that is technically easier than previous approaches. However, like every tomography method, it requires a large number of tomograms. Below, we show in detail that, if the goal is not full mechanical tomography, but merely demonstrating mechanical nonclassicality, this can be done much more simply using a single $s$-parameterized mechanical tomogram.

In the following we focus on the nonclassicality signified by negativities of the P-function, i.e.\ the quasiprobability distribution for $s=1$, as given in Eq.~\eqref{eq:Pfunc}. This is a favourable choice over the Wigner function, since P-negativity is more resilient to noise and loss, and captures some non-classical states, such as squeezed vacuum states, which have a positive Wigner function. Furthermore, P-function negativity can, in principle, be directly observed using appropriate filtering of the homodyne data~\citep{Kiesel2008,Kiesel2011,Agudelo2015}, or \emph{witnessed} without full tomography, using only a small number of observables~\citep{Shchukin2005,Shchukin2005-2,Kiesel2009,Kiesel2012,Ryl2015}.

\begin{figure}[t]
  \includegraphics[width=\columnwidth]{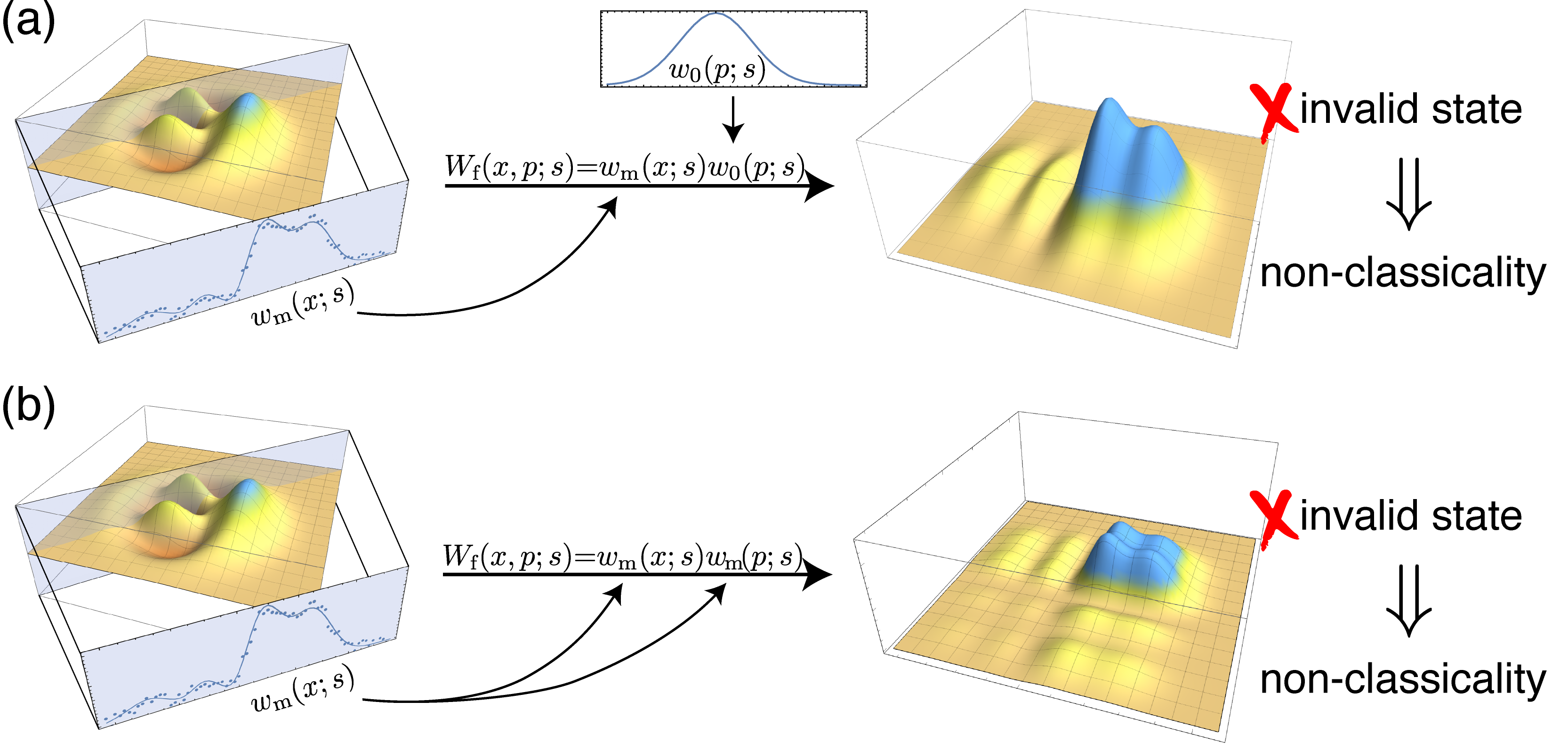}
  \caption{\textbf{Witnessing P-function non-classicality.} Using the method outlined in the main text (c.f.\ Fig.~\ref{fig:Tomography}) one can obtain a $s$-tomogram of the state of the mechanical resonator. Using this $s$-tomogram, one computes \textbf{(a)} the fictitious $s$-parameterized quasiprobability distribution $\mathscr{W}\s{f}(x,p;s) = w\s{m}(x;s)w_0(p;s)$, where $w_0(p;s)$ is an $s$-tomogram of the vacuum state, or \textbf{(b)} the fictitious distribution $\mathscr{W}\s{f}(x,p;s) = w\s{m}(x;s)w\s{m}(p;s)$, using the same tomogram for both quadratures. If either of the the resulting functions does not correspond to a bona-fide density operator, then the state of the mechanical resonator has been P-nonclassical.}
  \label{fig:Witness}
\end{figure}

Unfortunately, these methods do not translate directly to our optomechanical scenario, where we only have access to a restricted set of observables, namely, tomograms. Nonetheless, building on a criterion introduced by Park~\textit{et al}~\citep{Park2017}, a single mechanical tomogram can be used to witness nonclassicality as follows.
Given a Wigner tomogram $w\s{m}(x)$ of a quantum state for some quadrature angle $\phi$ (here, for brevity, we have dropped the phase from the argument of the tomograms), one can construct a ficticious Wigner function (in the position-momentum basis) using the so-called \emph{first demarginalization} maps $w\s{m}(x) \mapsto \mathscr{W}\s{f}(x,p) = w\s{m}(x)w_0(p)$, where $w_0(p)=\sqrt{2}\exp\{-p^2\}$ is a marginal of the vacuum state.
If the fictitious Wigner function $\mathscr{W}\s{f}(x,p)$ is not a legitimate quantum state, then the state of the mechanical resonator has been P-nonclassical, see Fig~\ref{fig:Witness}a.

In case the first demarginalization cannot certify nonclassicality and the results are inconclusive, one can consider the \emph{second} demarginalization map which uses a duplication of a single tomogram by changing $x \rightarrow p$ in $w\s{m}(x)$, to construct a ficticious Wigner functions as $(w\s{m}(x),w\s{m}(x)) \mapsto \mathscr{W}\s{f}(x,p) = w\s{m}(x)w\s{m}(p)$.
Similarly, the failure of $\mathscr{W}\s{f}(x,p)$ to represent a legitimate quantum state implies the P-nonclassicality of the state of the mechanical resonator, see Fig~\ref{fig:Witness}b. Importantly, testing whether the result is a legitimate Wigner function can be relatively simple in many cases of interest~\citep{Park2017}. These simple, necessary but not sufficient, criteria are thereby able to detect the nonclassicality of a large class of quantum states, including phonon added thermal states, which are of particular current interest.

As we show in detail in the Appendix~\ref{app:sNonclass}, this procedure can be directly extended to any $s$-parametrized quasiprobability distribution. Given an $s$-parameterized tomogram of a quantum state, $w\s{m}(x;s)$, one can define the following two demarginalization maps,
\begin{align}
&(w\s{m}(x;s),w_0(x;s)) \mapsto \mathscr{W}\s{f}(x,p;s) = w\s{m}(x;s)w_0(p;s),\label{eq:sfic1}\\
&(w\s{m}(x;s),w\s{m}(x;s)) \mapsto \mathscr{W}\s{f}(x,p;s) = w\s{m}(x;s)w\s{m}(p;s)\label{eq:sfic2},
\end{align}
in which $w_0(x;s)$ is the $s$-parameterized tomogram of the vacuum state and $\mathscr{W}\s{f}(x,p;s)$ is the $s$-parameterized quasiprobability distribution of a fictitious quantum state. The failure of either of the distributions in Eq.~\eqref{eq:sfic1} or~\eqref{eq:sfic2} to represent a bona fide quantum state implies the P-function nonclassicality of the corresponding quantum state. In other words, the two maps presented above provide necessary nonclassicality criteria.

To verify the legitimacy of the quantum state corresponding to the above fictitious quasiprobability distributions, one can simply employ all the arguments provided in Ref.~\citep{Park2017} by using appropriate quasiprobability distributions to compute the matrix elements of the fictitious density operator using the phase-space trace relation. This shows that our method can be used to verify P-function nonclassicality even in the regime of weak interactions.

\subsubsection{Effect of imperfections}\label{sec:imperfections}

An important question that remains is whether experimental noise and losses could lead to a false positive result in our proposed nonclassicality-verification procedure. Since any noise on the mechanical resonator can be considered part of the mechanical quantum state, we only have to study the effect of noise and losses on the optical readout signal.

Considering first the readout channel before the optomechanical interaction, note that the optical input in Eq.~\eqref{eq:SVS} has a Gaussian Wigner function. Since displacement operations as well as all experimental noise and losses are represented by Gaussian channels~\citep{Eisert2005}, the state always remains Gaussian and can easily be fully characterised; see for example~\citep{Parthasarathy2015}. One can then use this experimentally characterized state in Eq.~\eqref{eq:tomomain1} and proceed as before. Considering now the noise and losses after the interaction, we show in Appendix~\ref{app:Noise} that these commute with the interaction and can thus be treated in the same way.

Given that these imperfections are well characterized, they can then be taken into account by correcting the parameter $s$ of the fictitious tomogram $w_0(p;s)$ in Eq.~\eqref{eq:sfic1}. Curiously, this is not necessary for our second nonclassicality criterion, Eq.~\eqref{eq:sfic2}, which is immune to noise and losses by virtue of using the same measured tomogram for both quadratures. The latter implies that both tomograms have matching order parameters.
Nonetheless, since the legitimacy tests in Appendix~\ref{app:sNonclass}, require using the $s$-parameterized phase-space trace rule, a good estimate of the experimental $s$ is necessary for both criteria. Notably, a benchmarking of the error is always possible experimentally~\citep{Park2017} by preparing classical states of the mechanics, e.g.\ thermal states at different temperatures. As a result, the reliability of our nonclassicality criteria is not affected by the reduction in the experimentally achievable $s$ due to noise and loss effects. Moreover, since the demarginalization maps of Eqs.~\eqref{eq:sfic1} and~\eqref{eq:sfic2} are informationally equivalent for all values of $s$, the above mentioned reduction does not decrease the power of our nonclassicality criteria.

We now comment briefly on the effect of displacement-amplitude fluctuations that cause random fluctuations in the interaction strength $\chi$ given by Eq.~\eqref{eq:chi}.
First, we note that both displacements would in practice be performed using the same laser and can thus be implemented using inherently stable designs, such as Sagnac or polarization-based interferometers.
Second, our scheme properly measures only one mechanical quadrature at a time, and it is well-known that a single quadrature of a harmonic oscillator is a continuous quantum non-demolition (QND) observable that can be measured precisely and continuously~\citep{Thorne1978}.
In our scenario, this is reflected in the fact that for any pulse length $\tau$ the mechanical bosonic annihilation operator $b'$ commutes with itself (see Eq.~\eqref{SI:eq:Totalaprime} in Appendix~\ref{app:Transes}), meaning that there is no back-action from light onto the mechanics in our measurement scheme~\citep{Braginsky1980}.
Third, in light of Eq.~\eqref{eq:GkernelTomo} it turns out that fluctuations in the amplitude and consequently $\chi$ only affect the width of the tomographic kernel, namely it corresponds to small fluctuations in the parameter $s^\star$ given by Eq.~\eqref{eq:IntOrder}.
Therefore, as long as fluctuations give rise to small relative errors in the ordering parameter, they can be safely ignored.
Finally, we emphasize here that amplitude fluctuations are not amplified when squeezing is exploited as it takes place before the displacement.

Besides the above analysis, it is important to consider the possibility of false positive (or false negative) results due to experimental noise and finite statistics. For this we first use the following Lemma from Ref.~\citep{Ferraro2012}:
\begin{lemma}\label{lemma2}
The set of P-classical states is nowhere dense in the whole bosonic state space~\citep{Ferraro2012}.
\end{lemma}
\noindent Lemma~\ref{lemma2} together with the fact that the set of P-classical states is closed implies that this set cannot have an accumulation point within the bosonic state space that is not P-classical.
The latter means that, for every P-nonclassical state there exists a sufficiently small epsilon-ball (in state space) which contains no P-classical states. In other words, given a P-nonclassical state, it will remain P-nonclassical under sufficiently small perturbations. As a consequence, our (or any) method for nonclassicality detection is robust to fluctuations and finite-statistics effects for false-positive results, as long as these are small enough.

In practice, the fluctuations might of course be larger, such that they could in principle lead to false conclusions. Similar problems are well-known from quantum state tomography, where statistical fluctuations might, for example, make a state look entangled, although it is separable. Such false positives can be identified using standard techniques such as Monte-Carlo resampling according to the known experimental error model. Such techniques are, for instance, well established in the context of discrete-variable quantum tomography and produce reliable confidence regions for the reconstructed quantum state. In these methods the finite-statistic problem is taken care of by choosing a sufficiently large number of samples, typically determined by the properties of the state of interest through Hoeffding's inequality.

On the other hand, given an initial nonclassical state, noisy measurements can wash out the nonclassical features and lead to a false-negative result. While this cannot be avoided in general, it should be noted that the arguments above imply that sufficiently small fluctuations will preserve the nonclassical features. However, the exact magnitude of noise and finite statistics that can be tolerated depends on the nonclassicality of the state under test and the conditions of the experiment.

\subsection{Experimental Proposal} \label{sec:experiment}
We now describe a practical protocol for performing tomography of a mechanical resonator using our technique.
First, recall that the phase $\varphi$ is set by the pulse duration $\tau$ via the condition of Eq.~\eqref{eq:cond} and substituting into Eq.~\eqref{eq:mecphase}.
For a given pulse duration and depending on the optomechanical system parameters, the best pulse amplitude is then $|\alpha|=r$ such that the energy-time uncertainty relation is satisfied. Although using these parameters fixes the readout quadrature to $\varphi=\varphi_0$, Eq.~\eqref{eq:Mtomogram} shows that arbitrary quadrature angles $\varphi\s{d}$ can be probed by exploiting the free evolution of the mechanics in the time delay $\tau\s{d}$ between state preparation and readout, which can be precisely controlled in practice, see Fig~\ref{fig:Tomography}a.
For each quadrature angle, one then measures the momentum tomogram of the Wigner function of the reflected light, from which the mechanical tomogram can be extracted by simple rescaling according to Eq.~\eqref{eq:Mtomogram}. Repeating this process for each required tomogram, on obtains enough information to fully reconstruct the Wigner function of the mechanical resonator.

We will now study the parameters of a few state-of-the-art optomechanical experiments to test the feasibility of our approach. For this analysis we will consider no squeezing in the initial optical state, i.e.\ $\varepsilon=0$, as squeezing remains a challenging task in practice. Furthermore, since the optical pulses required for all these systems are relatively long, it incurs no experimental complications to assume that the pulse length can be chosen close to the optimal $\tau\sim \pi/\omega_{mec}$, with a relative stability on the order of $10^{-4}$, which, using Eqs.~\eqref{eq:mecphase} and~\eqref{eq:cond}, implies $u\sim 2$ and $\chi\sim \sqrt{k/2\sqrt{3}}$ for $k\in 2\mathbb{Z}^+$. In practice, these pulses would be generated by chopping continuous-wave lasers and can thus not only be timed very precisely, but also inherit the bandwidth and coherence properties of the parent laser enabling displacement-counter-displacement operations (as shown in Fig.~\ref{fig:Tomography}a) to high accuracy. We will study the more difficult task of reconstructing the mechanical Wigner function, rather than other $s$-parameterized quasiprobability distributions. We thus compute the requirements for $\chi\sim 3$ to satisfy the condition of Eq.~\eqref{eq:orderApprox}, up to the order of $10^{-3}$, which is below other typical noise sources in the experiment. Note that, Eq.~\eqref{eq:cond} also requires a relative stability of the laser power on the order of $10^{-3}$.

\setlength\tabcolsep{0.25em}
\begin{table*}[t]
\centering
\small{
\begin{tabular}{c | c c c c c c | c c }
Reference   &   $\omega_\text{m}$ [Hz]   &   mass [kg]   &   $\Gamma_\text{m}$ [Hz]   &   $g_0$ [Hz]   &   $\kappa_\text{o}$ [Hz]  &   $\kappa\s{o}/\omega_\text{m}$   &   opt $\tau$ [s]   &  pulse Energy [J] \\ 
\hline\hline
Kleckner \textit{et al}~\citep{Kleckner2011}     &     $9.7\times 10^3$     &     $1.1\times 10^{-10}$     &     $1.3 \times 10^{-2}$     &     $22$     &     $4.7\times 10^{5}$     &     55     &    $\sim 2.1\times 10^{-4}$    &    $\sim 5.5\times 10^{-13}$ \\
Murch \textit{et al}~\citep{Murch2008}     &     $4.2\times 10^4$     &     $10^{-22}$     &     $10^3$     &     $6\times 10^5$     &     $6.6\times 10^{5}$     &     15.7      &     $\sim 4.8\times 10^{-5}$    &    $\sim 1.4\times 10^{-20}$ \\
Norte \textit{et al}~\citep{Norte2016}     &     $1.5\times 10^5$     &     $10^{-12}$     &     $1.4 \times 10^{-3}$     &     $10^{2}$     &     $10^{6}$    &     6     &     $\sim1.3\times 10^{-5}$     &     $\sim 6.4\times 10^{-12}$\\
Thompson \textit{et al}~\citep{Thompson2007}     &     $1.3\times 10^5$     &     $4\times 10^{-11}$     &     $0.12$     &     $50$     &     $5\times 10^{5}$     &     3.7    &     $\sim 1.5\times 10^{-5}$    &     $\sim 1.9\times 10^{-11}$ \\
Anguiano \textit{et al}~\citep{Anguiano2016}     &     $20\times 10^9$     &     $7.7\times 10^{-12}$     &     $2 \times 10^6$     &     $4.8 \times10^7$     &     $3.4\times 10^{10}$     &     1.72     &     $\sim10^{-10}$    & $\sim 4.9\times 10^{-13}$ \\
Arcizet \textit{et al}~\citep{Arcizet2006}     &     $8.1\times 10^5$     &     $1.9\times 10^{-7}$     &     $81$     &     $1.2$     &     $10^{6}$     &     1.3     &     $\sim 2.5\times 10^{-6}$    &     $\sim1.3\times 10^{-6}$ \\
Cuthbertson \textit{et al}~\citep{Cuthbertson1996}     &     $10^3$     &     $1.85$     &     $2.5 \times 10^{-6}$     &     $1.2\times 10^{-3}$     &     275     &     0.9     &     $\sim2\times 10^{-3}$    &     $\sim 2.0\times 10^{-6}$ \\
Gr\"oblacher \textit{et al}~\citep{Groblacher2009}     &     $9.5\times 10^5$     &     $1.4\times 10^{-10}$     &     $1.4 \times 10^{2}$     &    $3.9$     &     $2\times 10^{5}$     &     0.22     &    $\sim2.1\times 10^{-6}$     &     $\sim1.7\times 10^{-7}$ \\
Chan \textit{et al}~\citep{Chan2011}     &     $3.9\times 10^9$     &     $3.1\times 10^{-16}$     &     $3.9 \times 10^3$     &     $9\times 10^5$     &     $5\times 10^{8}$     &     0.13     &     $\sim 5.1\times 10^{-10}$    &    $\sim 5.3\times 10^{-11}$ \\
Verhagen \textit{et al}~\citep{Verhagen2012}     &     $7.8\times 10^7$     &     $1.9\times 10^{-12}$     &     $3.4 \times 10^3$     &     $3.4\times 10^3$     &     $7.1\times 10^{6}$     &     0.09     &    $\sim 2.6\times 10^{-8}$     &     $\sim 1.5\times 10^{-9}$ \\
Teufel \textit{et al}~\citep{Teufel2011}     &     $1.1\times 10^7$     &     $4.8\times 10^{-14}$     &     $32$     &     $2\times 10^2$     &     $2\times 10^5$     &     0.02     &     $\sim 1.8\times 10^{-7}$    &     $\sim 8.5\times 10^{-9}$ 
\end{tabular}}
\caption{Experimental parameters for different optomechanical systems. For each system we provide the optimal pulse length $\tau$ for the probe pulse used by our scheme, together with the required pulse energy for Wigner-function tomography based on the parameter assumptions $\varepsilon=0$, $u=2$, and $\chi\sim 3$.}
\label{tab:systems}
\end{table*}

The relevant parameters for a number of optomechanical systems, taken mainly from Ref.~\citep{Aspelmeyer2014} are summarized in Tab.~\ref{tab:systems}. We note that although many of these systems are indeed in the resolved sideband regime, they could nonetheless employ our technique by sufficiently decreasing the quality factor of the used optical cavities, as outlined in Tab.~\ref{tab:systems}. Let us now consider explicitly kHz-frequency resonators such as $Si_3N_4$ membranes~\citep{Kleckner2011,Norte2016}, which are promising candidates for ground-state cooling~\citep{Groblacher2009} and state preparation via measurement-based methods~\citep{Ringbauer2016Optomechanics}. In this case, the required pulse duration is on the order of tens of microseconds, which can easily be achieved, even with gain-switched CW-lasers.
In view of Eqs.~\eqref{eq:chi} and~\eqref{eq:cond}, the requirements on the optical power to reach $\chi\sim3$ then depends on the optomechanical coupling strength $g_0$ for the respective system and the readout displacement amplitude. In practice, optical powers on the order of mW or below are sufficient for almost all systems presented in Tab.~\ref{tab:systems}. Evidently, it must be ensured that these powers are tolerated by the corresponding mechanical resonators, which are typically operated at lower optical powers. Should it be necessary to decrease the optical power impinging on the resonator to prevent damaging it, one could consider using a train of readout pulses with a repetition rate equal to the mechanical frequency and a total duration much shorter than the mechanical decay time. This approach greatly decreases the required optical power without incurring significant added noise.

Note, that the input power requirements of our method can be decreased significantly by using a low finesse optical cavity, such that the optical decay rate $\kappa\s{o}$ is much larger than the mechanical frequency $\omega\s{m}$. As an example, the system of Ref.~\citep{Kleckner2011} in Tab.~\ref{tab:systems} has a suitable cavity with a sideband resolution factor of $\kappa\s{o}/\omega\s{mec}\sim 50$, and is thus in the ideal regime for our method. Making use of the optical cavity with a finesse of $\sim 3\times 10^4$, the required optical power is on the order of $10^{-13}$ W. Optical displacements~\citep{Becerra2013} and homodyne tomography of the optical output field can then be performed using standard methods~\citep{Lvovsky2009,Barbieri2010}.

Finally, we would like to point out that the use of squeezed readout light, though not necessary, would be highly beneficial for our protocol through reducing the requirements on amplitude and precision of the optical displacement operations. Specifically, one could choose $\chi \sim 1$ which, in turn, requires a squeezing parameter $\varepsilon \sim \ln(3)$ to reach the same approximation as before.  This is equivalent to a vacuum squeezing of $\sim 9.5$dB, which is achievable with current technology.
In this case, both the requirements of Eqs.~\eqref{eq:cond} and~\eqref{eq:orderApprox} are satisfied easily without high precision in the control of the laser power.
Importantly, we note that squeezing does not affect the required stability in the laser power, as it takes place before the displacement. Furthermore, the squeezing parameter does not enter into the interaction transformation of Eq.~\eqref{eq:aprime}, implying that it need not even be precisely characterized, as long as Eq.~\eqref{eq:orderApprox} is satisfied.

\section{Discussion} \label{sec:conclusion}
We have introduced a method that allows for the tomographic reconstruction of the phase space quasiprobability distributions of a mechanical resonator in a regime where the commonly used rotating-wave approximation does not hold. In contrast to conventional methods in which motional sidebands are resolved with the help of a high quality optical cavity, our approach fits within the bad or no cavity regime. As a result, our technique is applicable to a wide range of experiments. Taking into account the full optomechanical Hamiltonian, we demonstrated that, even in this regime, the mechanical quadrature distributions can be imprinted onto an optical field through the radiation pressure interaction. Carefully controlling for back-action, optical displacements, and the optical Kerr effect, we showed how this information can be extracted using established methods and current technology. Finally, we discussed how our approach could be used to witness nonclassicality of the mechanical state in a much more resource-efficient way without requiring full quantum state reconstruction. We anticipate that our tomographic technique will enable the reconstruction of the quantum states of motion of mechanical resonators, and verify their potential nonclassicality, in a wide range of experimental platforms, without the challenging requirements of alternative approaches.

\begin{acknowledgments}
We thank M.\ Vanner, A.G.\ White, W.\ Bowen, and A.\ Szorkovszky for helpful discussions. 
This work was supported in part by the ARC Centres of Excellence for Engineered Quantum Systems (CE110001013, CE170100009) and Quantum Computation and Communication Technology (CE110001027), and the UK Engineering and Physical Sciences Research Council (grant number EP/N002962/1).
F.S. developed the concepts. F.S. and M.R. worked out the calculations and experimental parameters. All authors wrote the manuscript.
\end{acknowledgments}

\newpage
~
\newpage

\section{Methods}\label{sec:methods}
\subsection{Phase-space representation} \label{sec:PSRep}
Bosonic continuous-variable quantum systems are best described using a quantum phase-space representation with respect to two orthogonal quadratures, such as position and momentum. For such a representation one requires an informationally-complete set of basis operators, which are commonly taken to be the Weyl-Wigner operators $\hat{T}(\alpha;s) = \dmu\xi \exp\{\alpha\xi^*-\alpha^*\xi + \frac{s}{2}|\xi|^2\} \hat{D}(\xi)$, for a complex-valued phase-space point $\alpha\in\mathbb{C}$ and distribution parameter $s\in[-1,1]$~\citep{GlauberBook}. Here, $\hat{D}(\xi){=}\exp\{\xi a^\dag - \xi^* a\}$ is the displacement operator by the value $\xi\in\mathbb{C}$, and $a$ ($a^\dag$) is the bosonic annihilation (creation) operator. Weyl-Wigner operators are normal, complete, and satisfy the orthogonality (duality) relation,
\begin{align}
& {\rm Tr}\hat{T}(\alpha;s){=}1, \label{eq:Tnorm}\\
& \dmu\alpha \hat{T}(\alpha;s) = \hat{I}, \label{eq:Tcomp}\\
& {\rm Tr}\hat{T}(\alpha;s)\hat{T}(\beta;-s){=}\pi\delta^{(2)}(\alpha{-}\beta), 
\label{eq:Torth}
\end{align}
respectively.
As a result, we can expand any operator $\hat{\Lambda}$ in this basis as
\begin{equation}\label{eq:Wrep}
\hat{\Lambda} = \dmu\alpha \mathscr{W}_{\hat{\Lambda}}(\alpha;s) \hat{T}(\alpha;-s).
\end{equation}
The distribution $\mathscr{W}_{\hat{\Lambda}}(\alpha;s)={\rm Tr}\hat{\Lambda}\hat{T}(\alpha;s)$ is known as the $s$-parameterized quasiprobability distribution of the operator $\hat{\Lambda}$, which includes as special cases~\citep{GlauberBook} the Husimi-Kano Q-function~\citep{Husimi1940,Kano1965} for $s=-1$, the Wigner~\citep{Wigner1932} function for $s=0$, and the Glauber-Sudarshan P-function~\citep{Glauber1963,Sudarshan1963} for $s=1$. Crucially, these distributions, which are easily generalized to the multipartite case~\citep{VogelBook}, are not, in general, legitimate probability distributions, as they can take negative values. If the operator $\hat{\Lambda}$ is a quantum state, then such negative values are the primary signature of nonclassical behaviour~\citep{VogelBook,Sperling2016}, and can always be detected using direct observation via tomographic methods~\citep{Leonhardt1997,Kiesel2008,Kiesel2011,Agudelo2015} or nonclassicality criteria~\citep{Vogel2000,Shchukin2005,Shchukin2005-2,Kiesel2009,Kiesel2012,Ryl2015,Sperling2017}. 

Using Eq.~\eqref{eq:Wrep} and the duality relation of Eq.~\eqref{eq:Torth}, the trace operation can be expressed in terms of phase-space distributions as
\begin{equation}\label{eq:trace}
{\rm Tr}\hat{\Lambda}\hat{\Upsilon} = \dmu\alpha \mathscr{W}_{\hat{\Lambda}}(\alpha;s)\mathscr{W}_{\hat{\Upsilon}}(\alpha;-s) .
\end{equation}
Finally, it is sometimes more natural to work with position-momentum coordinates, in which case $d^2\alpha = d{\rm Re}\alpha\, d{\rm Im}\alpha=\frac{dqdp}{2}$, and $\mathscr{W}_{\hat{\Lambda}}(\alpha;s)=2\pi\mathscr{W}_{\hat{\Lambda}}(q,p;s)$, so that
\begin{equation}
\begin{split}
\dmu\alpha \mathscr{W}_{\hat{\Lambda}}(\alpha;s) = \int dqdp \mathscr{W}_{\hat{\Lambda}}(q,p;s) = 1.
\end{split}
\end{equation}

\subsection{Phase-space tomography}\label{sec:PSTomo}
\begin{figure}[t]
  \includegraphics[width=\columnwidth]{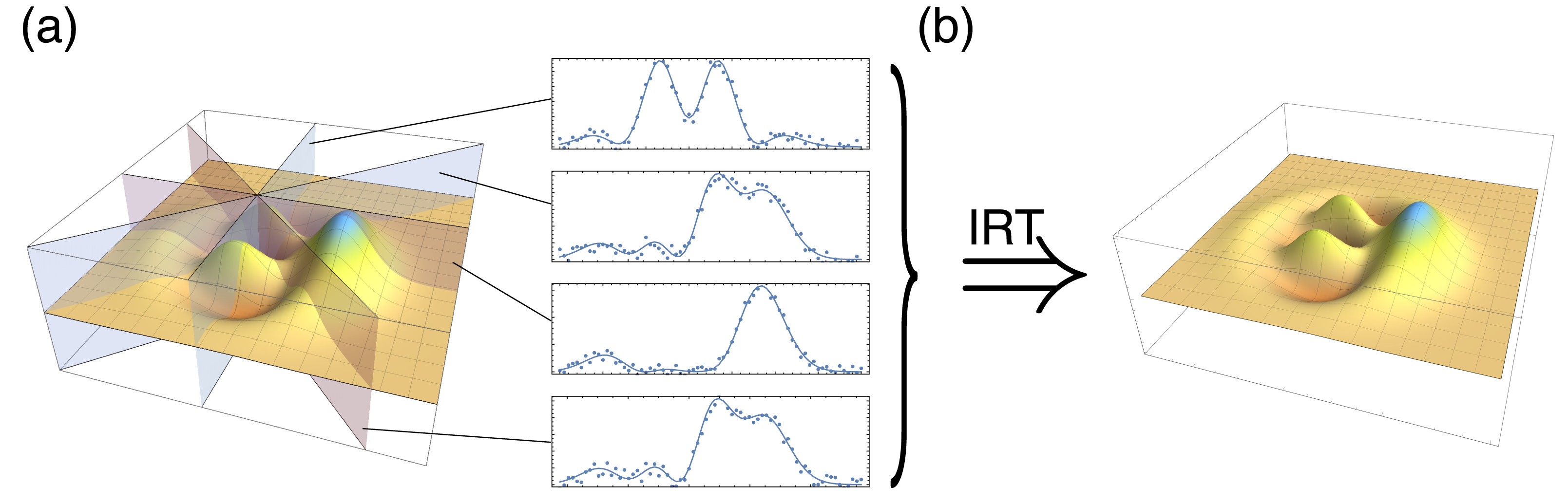}
  \caption{\textbf{Traditional Wigner-function tomography.} \textbf{(a)} The continuous-variable quantum system, in an initially unknown quantum state, is subject to quadrature measurements, for example, through homodyne detection. \textbf{(b)} Using a sufficiently large number of quadrature measurements (depending on the complexity of the unknown state), an inverse Radon transformation (IRT) can be used to reconstruct the Wigner function of the unknown state.}
  \label{fig:Motivation}
\end{figure}

\begin{figure}[b!]
  \includegraphics[width=\columnwidth]{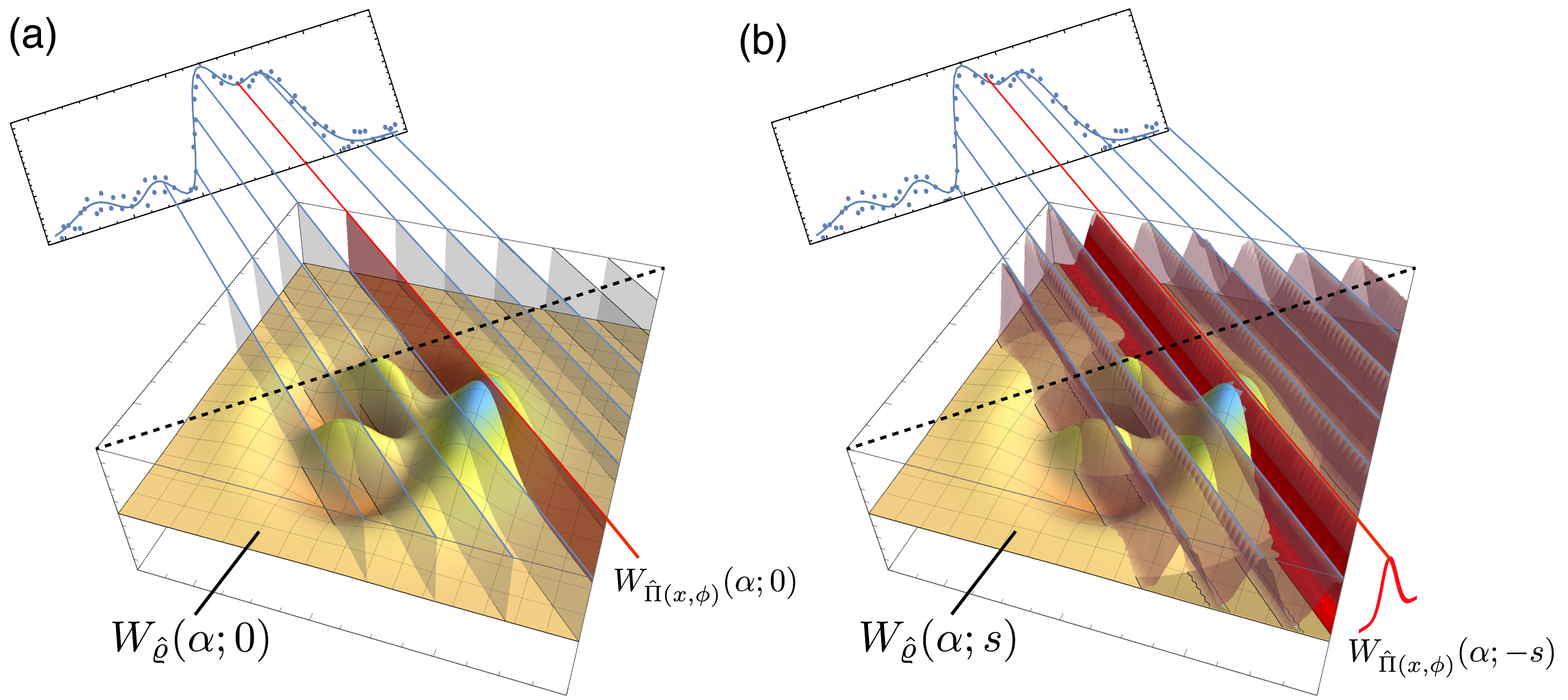}
  \caption{\textbf{Radon transform over phase space.} \textbf{(a)} We can think of the standard Radon transform of Eq.~\eqref{eq:radon} as sampling Wigner function of a quantum system along different quadrature axes via a string of zero width (a Dirac delta function cross section). \textbf{(b)} In the case of Eq.~\eqref{eq:sradon}, one samples an $s$-parametrized quasiprobability distribution with a string of finite width, that is, a Gaussian cross section.}
  \label{fig:sampling}
\end{figure}

While the quasiprobability distribution of an unknown quantum state cannot be measured directly, it can be tomographically reconstructed using a set of experimentally accessible marginal quadrature distributions or \emph{tomograms}. Recall that the $\phi$-quadrature of a bosonic field is given by $\hat{x}(\phi) = \left(a e^{-i\phi} + a^\dag e^{i\phi}\right)/\sqrt{2}$ with eigenvalues and eigenvectors defined by the eigenvalue equation $\hat{x}(\phi)\ket{x(\phi)} = x(\phi) \ket{x(\phi)}$.
The set of all projections onto quadrature eigenstates $\{\hat{\Pi}(x,\phi)=\ket{x(\phi)}\bra{x(\phi)}\}$ for $x\in\mathbb{R}$ and $\phi\in [0,2\pi)$, known as \emph{tomographic operators}, forms an informationally complete operator basis~\citep{Leonhardt1997,Schleich2015}. One can then define the set of tomograms $\{w(x,\phi)\}$ for a quantum state $\hat{\varrho}$ as
\begin{equation}\label{eq:tomogram}
w(x,\phi) = {\rm Tr}\hat{\varrho} \hat{\Pi}(x,\phi) = \bra{x(\phi)} \hat{\varrho} \ket{x(\phi)} ,
\end{equation}
Using Eq.~\eqref{eq:trace}, this tomographic relation can be expressed in terms of Wigner functions as,
\begin{equation}\label{eq:radon}
w(x,\phi) = \dmu\alpha \mathscr{W}_{\hat{\varrho}}(\alpha;0)\mathscr{W}_{\hat{\Pi}(x,\phi)}(\alpha;0) ,
\end{equation} 
where $\mathscr{W}_{\hat{\varrho}}(\alpha;0)$ is the Wigner function of the quantum state, and
\begin{equation}\label{eq:tomOpWigner}
\mathscr{W}_{\hat{\Pi}(x,\phi)}(\alpha;0) = \delta(x-\frac{e^{i\phi}\alpha^* + e^{-i\phi}\alpha}{\sqrt{2}})
\end{equation}
is the Wigner function of the tomographic operator $\hat{\Pi}(x,\phi)$. Equation~\eqref{eq:radon}, also known as a \emph{Radon transformation}, can be interpreted as sampling the Wigner function of a quantum state $\hat{\varrho}$ with strings of zero width, see Fig.~\ref{fig:sampling}a. In turn, a sufficiently large set of tomograms can be used to reconstruct the Wigner function via an \emph{inverse Radon transformation}, see Fig.~\ref{fig:Motivation}.
In short,
\begin{equation}
\begin{split}
&\mathscr{W}_{\hat{\varrho}}(\alpha;0) \quad \xmapsto[{\rm Tr}\hat{\varrho} \hat{\Pi}(x,\phi)]{\text{Radon transf.}}\quad \{w(x,\phi)\},\\
&\{w(x,\phi)\} \quad \xmapsto{\text{Inverse Radon transf.}}\quad \mathscr{W}_{\hat{\varrho}}(\alpha;0).
\end{split}
\end{equation}
The tomographic relation~\eqref{eq:tomogram} can be generalized to arbitrary phase-space quasiprobability distributions with $s\geqslant 0$ as~\citep{Bazrafkan2011}
\begin{equation}\label{eq:sradon}
w(x,\phi) = \dmu\alpha \mathscr{W}_{\hat{\varrho}}(\alpha;s)\mathscr{W}_{\hat{\Pi}(x,\phi)}(\alpha;-s) ,
\end{equation} 
with the tomographic operator
\begin{equation}\label{eq:tomOps}
\mathscr{W}_{\hat{\Pi}(x,\phi)}(\alpha;-s)= \frac{1}{\sqrt{s\pi}} \exp\{ -\frac{(x-\frac{e^{i\phi}\alpha^* + e^{-i\phi}\alpha}{\sqrt{2}})^2}{s} \} .
\end{equation}
In analogy with the Radon transform, Eq.~\eqref{eq:sradon} can be interpreted as sampling an $s$-parameterized phase-space quasiprobability distribution with a string with Gaussian cross section of finite width, see Fig.~\ref{fig:sampling}b. Notably, Eqs.~\eqref{eq:radon} and~\eqref{eq:sradon} connect the \emph{same} experimentally measured tomograms $w(x,\phi)$ to different quasiprobability representations of the underlying quantum state.


\begin{widetext}
\appendix
\newpage
\section*{\hspace{-0.8em}\huge{Appendices}}

\section{Disentangling the interaction}
\label{app:Disentangle}
The full unitary interaction can be written as
\begin{equation}
\label{SI:eq:fstateOrig1}
\begin{split}
&\hat{D}(-\alpha)e^{-i \tau_0 \omega\s{o}a^\dag a - i \tau_0 \omega\s{m} b^\dag b} e^{-i \tau  [\omega\s{o}a^\dag a + \omega\s{m}b^\dag b + g_0 a^\dag a(b^\dag+b)]}\hat{D}(\alpha) = \\
& \qquad\qquad\qquad\qquad \hat{D}(-\alpha)
e^{-i (\tau_0 + \tau)  \omega\s{o}a^\dag a - i \tau_0 \omega\s{m} b^\dag b} 
e^{-i \tau  [\omega\s{m}b^\dag b + g_0 a^\dag a(b^\dag+b)]}\hat{D}(\alpha)
\end{split}
\end{equation}
in which $\hat{D}(\alpha)=\exp\{\alpha a^\dag- \alpha^* a\}$ is the optical displacement operator, $\hat{\tau}$ is light pulse duration, and $\tau_0$ is the time between the optomechanical interaction and the optical displacement.
The r.h.s of Eq.~\eqref{SI:eq:fstateOrig1} follows from the fact that $a^\dag a$ commutes with all other terms of the Hamiltonian.
Now, to disentangle the mechanical free evolution during the optomechanical interaction, we use the disentangling relations (2.34) and (2.85)--(2.87) of Ref.~\citep{Puri2001}, namely
\begin{equation}
\begin{split}
\exp\{\theta [ \alpha_1 b^\dag b + \alpha_2 b + \alpha_3 b^\dag ]\} & =
\exp\{f_3 b^\dag \} \exp\{ f_1 b^\dag b \} \exp\{ f_2 b \}\exp\{f\s{x}(\theta)\}\\
& = \exp\{ f_1 b^\dag b \} \exp\{f_4 b^\dag \} \exp\{ f_2 b \}\exp\{f\s{x}(\theta)\}\\
& = \exp\{ f_1 b^\dag b \} \exp\{f_2 b + f_4 b^\dag \} \exp\{f\s{x}(\theta) - \frac{1}{2} f_2 f_4\}\\
& = \exp\{ f_1 b^\dag b \} \exp\{f_2 b + f_4 b^\dag \} \exp\{f_5\},
\end{split}
\end{equation}
where
\begin{equation}
\begin{split}
& f_1 = \alpha_1\theta,\qquad
f_2 = \frac{\alpha_2}{\alpha_1}(e^{\theta\alpha_1}-1), \qquad
f_3 = \frac{\alpha_3}{\alpha_1}(e^{\theta\alpha_1}-1), \qquad
f\s{x} = \frac{\alpha_2\alpha_3}{\alpha_1^2}(e^{\theta\alpha_1}-\theta\alpha_1 - 1),\\
& f_4 = f_3 e^{-f_1} = \frac{\alpha_3}{\alpha_1}(1 - e^{-\theta\alpha_1}), \qquad
f_5 = \frac{\alpha_2\alpha_3}{\alpha_1^2}(\sinh(\theta\alpha_1)-\theta\alpha_1).
\end{split}
\end{equation}
This gives
\begin{equation}
\label{SI:eq:fstateOrig2}
\begin{split}
&\hat{D}(-\alpha)
e^{-i \tau_0  \omega\s{o}a^\dag a}
e^{ - i \tau_0 \omega\s{m} b^\dag b} 
e^{-i \tau  [\omega\s{o}a^\dag a + \omega\s{m}b^\dag b + g_0 a^\dag a(b^\dag+b)]}\hat{D}(\alpha) =\\
& \hat{D}(-\alpha) 
e^{-i (\tau_0 + \tau)  \omega\s{o}a^\dag a } 
e^{-i (\tau + \tau_0)  \omega\s{m}b^\dag b} 
e^{\frac{g_0}{\omega\s{m}}a^\dag a[(1-e^{i\omega\s{m}\tau }) b^\dag + (e^{-i\omega\s{m}\tau }-1) b]} 
e^{\frac{g_0^2}{\omega\s{m}^2}(a^\dag a)^2[i\omega\s{m}\tau  - i\sin(\omega\s{m}\tau )]}\hat{D}(\alpha) = \\
& e^{-i (\tau + \tau_0)  \omega\s{m}b^\dag b} 
\hat{D}(-\alpha)
e^{-i (\tau_0 + \tau)  \omega\s{o}a^\dag a} 
e^{\frac{g_0^2}{\omega\s{mec}^2}(a^\dag a)^2[i\omega\s{m}\tau  - i\sin(\omega\s{m}\tau )]} 
e^{\frac{g_0}{\omega\s{m}}a^\dag a(\mu b^\dag - \mu^* b)}
\hat{D}(\alpha) =\\
& \underbrace{e^{-i (\tau + \tau_0)  \omega\s{m}b^\dag b}}
\underbrace{\hat{D}(-\alpha) e^{-i (\tau_0 + \tau)  \omega\s{o}a^\dag a} \hat{D}(\alpha)}
\underbrace{\hat{D}(-\alpha) e^{\frac{g_0^2}{\omega\s{m}^2}(a^\dag a)^2[i\omega\s{m}\tau  - i\sin(\omega\s{m}\tau )]} \hat{D}(\alpha)}
\underbrace{\hat{D}(-\alpha)  e^{\frac{g_0}{\omega\s{m}}a^\dag a(\mu b^\dag - \mu^* b)}
\hat{D}(\alpha)},\\
&\mu = 1-e^{i\omega\s{m}\tau },
\end{split}
\end{equation}
in which $\hat{D}(\alpha)=\exp\{\alpha a^\dag- \alpha^* a\}$ is the optical displacement operator. In Eq.~\eqref{SI:eq:fstateOrig2}, whenever needed we have used the fact that the optical and mechanical mode operators commute, and that $a^\dag a$ commutes with all other terms of the Hamiltonian.
In the resulting expression, we note that the first bracket is the free rotation of the mechanics in phase space, which can be eliminated by moving to a rotating frame with frequency $\omega\s{m}$. 

It is important to note that the optical state that we will be reconstructing contains all the evolutions described by the second, third, and fourth unitary evolutions.
Using the displacement relations
\begin{equation}\label{SI:eq:displacement}
\hat{D}(-\alpha)a\hat{D}(\alpha)=a+\alpha,\quad \hat{D}(-\alpha)a^\dag\hat{D}(\alpha)=a^\dag+\alpha^*,
\end{equation}
the second bracket is understood to be the free rotation of the optical field around a phase space point located at $-\alpha$.
The inverse of this operation can be easily applied to the reconstructed optical state and can thus be neglected for now.

The third bracket represents a bare nonlinear Kerr effect on the optical mode. The effect of this evolution can also be removed from the reconstructed optical state using numerical techniques. However, in the linearized regime we can take it into account as follows. Using Eq.~\eqref{SI:eq:displacement}, setting $\alpha = re^{i\theta}$, and noting the fact that $|\alpha|=r\gg 1$, we can approximate
\begin{equation}
\label{SI:eq:kerrapprox}
\begin{split}
\hat{D}(-\alpha) e^{\frac{g_0^2}{\omega\s{m}^2}(a^\dag a)^2[i\omega\s{m}\tau  - i\sin(\omega\s{m}\tau )]} \hat{D}(\alpha) & \simeq
e^{\frac{g_0^2 r^2}{\omega\s{m}^2}[4 a^\dag a + 2r(e^{i\theta}a^\dag + e^{-i\theta}a) + (e^{2i\theta} a^{\dag 2} + e^{-2i\theta}a^2) + r^2 + 1][i\omega\s{m}\tau  - i\sin(\omega\s{m}\tau )]}\\
& = e^{\chi^2 [4 a^\dag a + 2r(e^{i\theta}a^\dag + e^{-i\theta}a) + (e^{2i\theta} a^{\dag 2} + e^{-2i\theta}a^2)][i\omega\s{m}\tau  - i\sin(\omega\s{m}\tau )]},\\
\chi = \frac{g_0 r}{\omega\s{m}},
\end{split}
\end{equation}
where we have kept only the terms of order $r^2$ or higher, and have ignored the overall phases corresponding to $r^2+1$. We can also rewrite the last bracket of Eq.~\eqref{SI:eq:fstateOrig2} as
\begin{equation}
\label{SI:eq:fstateapprox}
e^{\frac{g_0}{\omega\s{m}} (a^\dag+\alpha^*)(a+\alpha) (\mu b^\dag - \mu^* b)},
\end{equation}
and linearise the exponent as
\begin{equation}
\label{SI:eq:LinHamilton}
\begin{split}
& \frac{g_0}{\omega\s{m}}(a^\dag+\alpha^*)(a+\alpha) (\mu b^\dag - \mu^* b)\\
& \qquad \qquad = \frac{g_0}{\omega\s{m}}(\alpha^* \mu a b^\dag - \alpha \mu^* a^\dag b)+\frac{g_0}{\omega\s{m}}(\alpha \mu a^\dag b^\dag - \alpha^* \mu^* ab) + \frac{g_0}{\omega\s{m}} a^\dag a(\mu b^\dag - \mu^* b) + \frac{g_0}{\omega\s{m}}r^2 (\mu b^\dag - \mu^* b)\\
& \qquad \qquad \simeq \frac{g_0ru}{\omega\s{m}} (e^{i\theta} a^\dag + e^{-i\theta} a + r)(e^{i\varphi} b^\dag - e^{-i\varphi} b),
\end{split}
\end{equation}
where $\mu = 1-e^{i\omega\s{m}\tau} = ue^{i\varphi}$ so that
\begin{equation}
\label{SI:eq:muparam}
\begin{split}
& u=|\mu|=\sqrt{2(1-\cos{\omega\s{m}\tau })},\\
& \varphi=\tan^{-1}{\left(\frac{\sin{\omega\s{m}\tau }}{1-\cos{\omega\s{m}\tau }}\right)} = \tan^{-1}{\left(\cot{(\omega\s{m}\tau/2)}\right)} = (m+1/2)\pi - \omega\s{m}\tau/2, \qquad m\in \mathbb{Z}^+ .
\end{split}
\end{equation}
Curiously, the radiation pressure interaction term $\frac{g_0}{\omega\s{m}} a^\dag a(\mu b^\dag+\mu^* b)$, has become negligible in Eq.~\eqref{SI:eq:LinHamilton}, since $\frac{g_0}{\omega\s{m}} \ll \frac{g_0ru}{\omega\s{m}}$ and can thus be omitted. Note that this also requires that $u$ be chosen large enough, which implies that the optimal pulse length $\tau$ is on the order of half the mechanical period. We can now define the new optical mode annihilation operator $\tilde{a}_\theta:=ae^{-i\theta}+\frac{r}{2}$ and the new mechanical annihilation operator $\tilde{b}_\varphi = b e^{-i\varphi}$ to get
\begin{equation}\label{SI:eq:intapprox}
e^{\frac{g_0}{\omega\s{m}} (a^\dag+\alpha^*)(a+\alpha) (\mu b^\dag - \mu^* b)} \simeq
e^{\chi u (\tilde{a}^\dag_\theta + \tilde{a}_\theta)(\tilde{b}^\dag_\varphi - \tilde{b}_\varphi)}.
\end{equation}
Note that, since $ 0 \leqslant u \leqslant 2$ and we work in a parameter regime where $\chi u=\frac{g_0ru}{\omega\s{m}}\sim 1-10$, it follows that $\chi^2 \geqslant \chi$ and hence the optical Kerr effect in Eq.~\eqref{SI:eq:kerrapprox} is not negligible compared to the interaction term of Eq.~\eqref{SI:eq:intapprox}.

Using the linearised evolutions of Eqs.~\eqref{SI:eq:kerrapprox} and~\eqref{SI:eq:intapprox}, we can now rewrite Eq.~\eqref{SI:eq:fstateOrig2} as
\begin{equation}
\label{SI:eq:approx}
\begin{split}
& \hat{\mathcal{U}}\s{o}~\hat{\mathcal{U}}\s{k}~\hat{\mathcal{U}}\s{om},\\
& \hat{\mathcal{U}}\s{om} = e^{\chi u(\tilde{a}^\dag_\theta + \tilde{a}_\theta)(\tilde{b}^\dag_\varphi - \tilde{b}_\varphi)},\\
& \hat{\mathcal{U}}\s{k} = e^{\chi^2[4 a^\dag a + 2r(e^{i\theta}a^\dag + e^{-i\theta}a) + (e^{2i\theta} a^{\dag 2} + e^{-2i\theta}a^2)][i\omega\s{m}\tau - i\sin(\omega\s{m}\tau)]},\\
& \hat{\mathcal{U}}\s{o} = \hat{D}(-\alpha) e^{-i (\tau_0 + \tau)  \omega\s{o}a^\dag a} \hat{D}(\alpha).
\end{split}
\end{equation}

\section{Transformation of the bosonic operators} \label{app:Transes}

\subsection{Optomechanical interaction}\label{app:OMint}

As outlined in the main text, we can now use Eq.~\eqref{SI:eq:approx} to compute the transformation of the vector of bosonic operators $\hat{\vec{A}}:=(a^\dag, b^\dag, a, b)$ under the optomechanical interaction given by 
\begin{equation} \label{SI:eq:OMAprime}
\hat{\vec{A}}'=\hat{\mathcal{U}}\s{om}\hat{\vec{A}}\hat{\mathcal{U}}^\dag\s{om}.
\end{equation}
Exploiting the properties of symplectic transformations~\citep{Wang1994}, we know that if $\hat{\mathcal{U}}=\exp(\frac{1}{2}\hat{\vec{A}} \ln M \Sigma \hat{\vec{A}}^\mathsf{T} )$, with $\Sigma=\begin{pmatrix}
 0 & I\\ 
 -I& 0
\end{pmatrix}$,
then $\hat{\mathcal{U}}\hat{\vec{A}}\hat{\mathcal{U}}^\dag= \hat{\vec{A}} M$.
In terms of the new the new field operators $\tilde{a}_\theta$ and $\tilde{b}_\varphi$,
\begin{equation}
\begin{split}
&\hat{\mathcal{U}}\s{om} = 
\exp\{ \frac{\chi u }{2} 
\begin{pmatrix}
\tilde{a}^\dag_\theta & \tilde{b}^\dag_\varphi & \tilde{a}_\theta & \tilde{b}_\varphi
\end{pmatrix}
\begin{pmatrix}
 0 & -1 & 0 & -1 \\ 
 1 & 0 & -1 & 0 \\
 0 & -1 & 0 & -1\\
 -1 & 0 & 1 & 0 
\end{pmatrix}
\begin{pmatrix}
\tilde{a}_\theta \\
\tilde{b}_\varphi \\
-\tilde{a}^\dag_\theta\\
-\tilde{b}^\dag_\varphi
\end{pmatrix}
 \},
 \end{split}
\end{equation}
which gives
\begin{equation}
\ln M = \chi u  \begin{pmatrix}
 0 & -1 & 0 & - 1 \\ 
 1 & 0 & -1 & 0 \\
 0 & -1 & 0 & -1\\
 -1 & 0 & 1 & 0 
\end{pmatrix},
\end{equation}
and thus,
\begin{equation}\label{SI:eq:OMMmatrix}
\begin{split}
& M = \begin{pmatrix}
 1 & -\chi u  & 0 & - \chi u  \\ 
 \chi u  & 1 & -\chi u  & 0 \\
 0 & -\chi u  & 1 & -\chi u \\
 -\chi u  & 0 & \chi u  & 1 
\end{pmatrix},~
M^{-1}=\begin{pmatrix}
 1 & \chi u  & 0 &  \chi u  \\ 
 -\chi u  & 1 & \chi u  & 0 \\
 0 & \chi u  & 1 & \chi u \\
 \chi u  & 0 & -\chi u  & 1 
\end{pmatrix},
\end{split}
\end{equation}
with $\det M=1$.
Using the matrix of Eq.~\eqref{SI:eq:OMMmatrix}, we find
\begin{equation}
\label{SI:eq:UatU}
\begin{split}
& \hat{\mathcal{U}}\s{om}\tilde{a}^\dag_\theta \hat{\mathcal{U}}\s{om}^\dag = \tilde{a}^\dag_\theta + \chi u  (\tilde{b}^\dag_\varphi - \tilde{b}_\varphi) = a^\dag e^{i\theta} +\frac{r}{2} - i\sqrt{2}\chi u  \hat{P}(\varphi),\\
& \hat{\mathcal{U}}\s{om}\tilde{b}^\dag_\varphi\hat{\mathcal{U}}\s{om}^\dag = \tilde{b}^\dag_\varphi - \chi u  (\tilde{a}^\dag_\theta + \tilde{a}_\theta) = b^\dag e^{i\varphi} - \sqrt{2} \chi u  [\hat{x}(\theta)+r]\\
& \hat{\mathcal{U}}\s{om}\tilde{a}_\theta\hat{\mathcal{U}}\s{om}^\dag = \tilde{a}_\theta - \chi u  (\tilde{b}^\dag_\varphi - \tilde{b}_\varphi) = a e^{-i\theta} +\frac{r}{2} + i\sqrt{2}\chi u  \hat{P}(\varphi),\\
& \hat{\mathcal{U}}\s{om}\tilde{b}_\varphi\hat{\mathcal{U}}\s{om}^\dag = \tilde{b}_\varphi - \chi u  (\tilde{a}^\dag_\theta + \tilde{a}_\theta) = b e^{-i\varphi} - \sqrt{2} \chi u  [\hat{x}(\theta)+r],
\end{split}
\end{equation}
where $\hat{P}(\varphi)=\frac{1}{i\sqrt{2}}(be^{-i\varphi} - b^\dag e^{i\varphi}) = \frac{1}{\sqrt{2}}(be^{-i(\varphi+\frac{\pi}{2})} + b^\dag e^{i(\varphi+\frac{\pi}{2})})=\hat{X}(\varphi + \frac{\pi}{2})$ is the $(\varphi + \frac{\pi}{2})$-quadrature of the mechanics, and $\hat{x}(\theta) = \frac{1}{\sqrt{2}}(a^\dag e^{i\theta} + a e^{-i\theta})$ is the $\theta$-quadrature of the optical field.
Making use of Eq.~\eqref{SI:eq:OMAprime},
\begin{equation}
\begin{split}
\hat{\mathcal{U}}\s{om} \tilde{a}_\theta \hat{\mathcal{U}}^\dag\s{om} &= \hat{\mathcal{U}}\s{om} a \hat{\mathcal{U}}^\dag\s{om} e^{-i\theta} + \frac{r}{2} = a'e^{-i\theta} + \frac{r}{2},\\
\hat{\mathcal{U}}\s{om} \tilde{b}_\varphi \hat{\mathcal{U}}^\dag\s{om} & = \hat{\mathcal{U}}\s{om} b \hat{\mathcal{U}}^\dag\s{om} e^{-i\varphi} = b' e^{-i\varphi},
\end{split}
\end{equation}
and thus, by virtue of Eq.~\eqref{SI:eq:UatU}, we find the following relations similar to that of Eq.~\eqref{eq:aprime} in the main text,
\begin{equation}\label{SI:eq:aprime}
\begin{split}
& a'^\dag = a^\dag + i\sqrt{2}\chi u  e^{-i\theta} \hat{X}(\varphi + \frac{\pi}{2}), \\
& b'^\dag = b^\dag - \sqrt{2} \chi u  e^{-i\varphi} [\hat{x}(\theta)+r],\\
& a' = a - i\sqrt{2}\chi u  e^{i\theta} \hat{X}(\varphi + \frac{\pi}{2}),\\
& b' = b - \sqrt{2} \chi u  e^{i\varphi} [\hat{x}(\theta)+r].
\end{split}
\end{equation}

As a consequence of the properties of the symplectic group $M$ represents a canonical (not necessarily unitary) transformation. This can be easily verified by computing the bosonic commutation relations $[a'^\dag,a']=[b'^\dag,b']=1$ and $[a',b']=[a'^\dag,b'^\dag]=0$. Note that the last term of the linearised Hamiltonian, Eq.~\eqref{SI:eq:LinHamilton}, only causes a constant displacement of the mechanics and could thus be absorbed into an initial displacement $b\mapsto b-\sqrt{2}\chi u  r e^{i\varphi}$. Finally, the transformation corresponding to Eq.~\eqref{SI:eq:aprime} is given by
\begin{equation}\label{SI:eq:SMatrixOM}
\begin{split}
S &= \begin{pmatrix}
 1 & -\chi u  e^{i\theta}e^{-i\varphi} & 0 & -\chi u  e^{i\theta} e^{i\varphi} \\ 
 i\chi u e^{-i\theta} e^{i(\varphi+\frac{\pi}{2})} & 1 & -i\chi u e^{i\theta} e^{i(\varphi+\frac{\pi}{2})} & 0 \\
 0 & -\chi u  e^{-i\theta} e^{-i\varphi} & 1 & -\chi u  e^{-i\theta} e^{i\varphi}\\
 i\chi u e^{-i\theta} e^{-i(\varphi+\frac{\pi}{2})} & 0 & -i\chi u e^{i\theta} e^{-i(\varphi+\frac{\pi}{2})} & 1
\end{pmatrix}\\
&= \begin{pmatrix}
 1 & -\chi u  e^{i\theta}e^{-i\varphi} & 0 & -\chi u  e^{i\theta} e^{i\varphi} \\ 
 -\chi u  e^{-i\theta} e^{i\varphi} & 1 & \chi u  e^{i\theta} e^{i\varphi} & 0 \\
 0 & -\chi u  e^{-i\theta} e^{-i\varphi} & 1 & -\chi u  e^{-i\theta} e^{i\varphi}\\
 \chi u  e^{-i\theta} e^{-i\varphi} & 0 & -\chi u  e^{i\theta} e^{-i\varphi} & 1
\end{pmatrix},\\
S^{-1} &= \begin{pmatrix}
 1 & \chi u  e^{i\theta}e^{-i\varphi} & 0 & \chi u  e^{i\theta} e^{i\varphi} \\ 
 \chi u  e^{-i\theta} e^{i\varphi} & 1 & -\chi u  e^{i\theta} e^{i\varphi} & 0 \\
 0 & \chi u  e^{-i\theta} e^{-i\varphi} & 1 & \chi u  e^{-i\theta} e^{i\varphi}\\
 -\chi u  e^{-i\theta} e^{-i\varphi} & 0 & \chi u  e^{i\theta} e^{-i\varphi} & 1
\end{pmatrix},\\
\det{S} &= 1.
\end{split}
\end{equation}

\subsection{Kerr interaction}
Along the same lines we can now evaluate the effect of the Kerr interaction on the optical field.
Again, given the vector of bosonic operators $\hat{\vec{A}}:=(a^\dag, b^\dag, a, b)$, we evaluate the output mode operators
\begin{equation} \label{SI:eq:KAprime}
\hat{\vec{A}}'=\hat{\mathcal{U}}\s{k}\hat{\vec{A}}\hat{\mathcal{U}}^\dag\s{k}.
\end{equation}
To find the result, we first define the new mode operators~\citep{Zhang1994}
\begin{equation}\label{SI:eq:anewKerr}
\bar{a} = a e^{-i\theta} +\frac{r}{3}.
\end{equation}
Substituting Eq.~\eqref{SI:eq:anewKerr} into the exponent of $\hat{\mathcal{U}}\s{k}$ in Eq.~\eqref{SI:eq:approx}, we obtain
\begin{equation*}
4 a^\dag a + 2r(e^{i\theta}a^\dag + e^{-i\theta}a) + (e^{2i\theta} a^{\dag 2} + e^{-2i\theta}a^2) = 
4\bar{a}^\dag \bar{a} + \bar{a}^{\dag 2} + \bar{a}^2 - \frac{6}{9}r^2.
\end{equation*}
The $r^2$ term is just an overall phase factor which can be changed arbitrarily.
Therefore, in terms of the new the new field operators $\bar{a}$, we have
\begin{equation*}
\begin{split}
&\hat{\mathcal{U}}\s{k} = 
\exp\{ \chi^2[i\omega\s{m}\tau  - i\sin(\omega\s{m}\tau )]
\begin{pmatrix}
\bar{a}^\dag & \bar{a}
\end{pmatrix}
\begin{pmatrix}
 2 & -1 \\
 1 & -2 
\end{pmatrix}
\begin{pmatrix}
\bar{a} \\
-\bar{a}^\dag
\end{pmatrix}
 \},
 \end{split}
\end{equation*}
Using the same relation as in Sec.~\ref{app:OMint},
\begin{equation*}
\ln M = 2 i v \begin{pmatrix}
 2 & -1 \\
 1 & -2 
\end{pmatrix},
\end{equation*}
where $v = \chi^2[\omega\s{m}\tau  - \sin(\omega\s{m}\tau )]$, and thus,
\begin{equation}\label{SI:eq:KMmatrix}
\begin{split}
& M =  \begin{pmatrix}
\cos{\bar{v}} + \frac{2i}{\sqrt{3}} \sin{\bar{v}} & -\frac{i}{\sqrt{3}} \sin{\bar{v}} \\
\frac{i}{\sqrt{3}} \sin{\bar{v}} & \cos{\bar{v}} - \frac{2i}{\sqrt{3}} \sin{\bar{v}}
\end{pmatrix},\quad
M^{-1}=\begin{pmatrix}
\cos{\bar{v}} - \frac{2i}{\sqrt{3}} \sin{\bar{v}} & \frac{i}{\sqrt{3}} \sin{\bar{v}} \\
-\frac{i}{\sqrt{3}} \sin{\bar{v}} & \cos{\bar{\bar{v}}} + \frac{2i}{\sqrt{3}} \sin{\bar{v}}
\end{pmatrix},
\end{split}
\end{equation}
in which $\bar{v} = 2 \sqrt{3}v$, and we have .
Using the matrix of Eq.~\eqref{SI:eq:KMmatrix}, we find
\begin{equation}
\label{SI:eq:KUatU}
\begin{split}
\hat{\mathcal{U}}\s{k}\bar{a} \hat{\mathcal{U}}\s{k}^\dag & = - \frac{i}{\sqrt{3}}\bar{a}^\dag \sin{\bar{v}} +  \bar{a} [\cos{\bar{v}} - \frac{2i}{\sqrt{3}}\sin{\bar{v}}] \\
& = - \frac{i}{\sqrt{3}} a^\dag e^{i\theta} \sin{\bar{v}} + a e^{- i\theta} (\cos{\bar{v}} - \frac{2i}{\sqrt{3}}\sin{\bar{v}}) + (\frac{1}{3}\cos{\bar{v}} - \frac{i}{\sqrt{3}}\sin{\bar{v}})r .
\end{split}
\end{equation}
Next, we have,
\begin{equation}
\begin{split}
\hat{\mathcal{U}}\s{k} \bar{a} \hat{\mathcal{U}}^\dag\s{k} &= \hat{\mathcal{U}}\s{k} a \hat{\mathcal{U}}^\dag\s{k} e^{-i\theta} + \frac{r}{3} = a'e^{-i\theta} + \frac{r}{3}.
\end{split}
\end{equation}
Equating this with Eq.~\eqref{SI:eq:KUatU}, we find 
\begin{equation}\label{SI:eq:Ktrans}
a' = - \frac{i}{\sqrt{3}} a^\dag e^{2i\theta} \sin{\bar{v}} + a (\cos{\bar{v}} - \frac{2i}{\sqrt{3}}\sin{\bar{v}}) + e^{i\theta}(\frac{1}{3}\cos{\bar{v}} - \frac{i}{\sqrt{3}}\sin{\bar{v}} - \frac{1}{3}) r.
\end{equation}
The linear transformation corresponding to Eq.~\eqref{SI:eq:Ktrans} is thus given by
\begin{equation}
\hat{\vec{A}}' =  \hat{\vec{A}} 
\begin{pmatrix}
\cos{\bar{v}} + \frac{2i}{\sqrt{3}}\sin{\bar{v}} & - \frac{i}{\sqrt{3}} e^{2i\theta} \sin{\bar{v}} \\
\frac{i}{\sqrt{3}} e^{-2i\theta} \sin{\bar{v}} &  \cos{\bar{v}} - \frac{2i}{\sqrt{3}}\sin{\bar{v}}
\end{pmatrix}
 + \vec{D},
\end{equation}
where the displacement vector $\vec{D}$ is defined as 
\begin{equation}
\vec{D} = (D^* \quad D) = \frac{r}{3} \left(e^{i\theta}(\cos{\bar{v}} + i \sqrt{3}\sin{\bar{v}} - 1)~,  \quad e^{-i\theta}(\cos{\bar{v}} - i \sqrt{3}\sin{\bar{v}} - 1) \right).
\end{equation}

\subsection{Optical transformation}

We now consider the effect of the last unitary evolution in Eq.~\eqref{SI:eq:approx}, namely, $\hat{\mathcal{U}}\s{o} = \hat{D}(-\alpha) e^{-i (\tau_0 + \tau)  \omega\s{o} a^\dag a} \hat{D}(\alpha)$.
This time, it is very easy to find the corresponding transformation in the Wigner representation using Lemma~1 (see the next section~\ref{app:ProofLem} for the proof this lemma).
The action of the displacement operator on the mode operators is given by Eq.~\eqref{SI:eq:displacement}, while the action of the phase-rotation operator is well-known to be
\begin{equation}\label{SI:eq:rot}
e^{-i\Theta a^\dag a} a e^{i\Theta a^\dag a} = a e^{i\Theta}, \qquad e^{-i\Theta a^\dag a} a^\dag e^{i\Theta a^\dag a} = a^\dag e^{-i\Theta}.
\end{equation}
It is then straightforward to evaluate the following:
\begin{equation}\label{SI:eq:Oprime}
a' = \hat{\mathcal{U}}\s{o} a \hat{\mathcal{U}}\s{o}^\dag = a e^{i(\tau_0+\tau)\omega\s{o}} + (e^{i(\tau_0+\tau)\omega\s{o}} - 1) \alpha.
\end{equation}
This represents a rotation of the optical field in phase space by an angle of $(\tau_0+\tau)\omega\s{o}$ around the phase-space point $\alpha$.

\subsection{The total evolution}

To obtain the effect of the overall evolution operator as described in Eq.~\eqref{SI:eq:approx}, we simply combine the results of Eqs~\eqref{SI:eq:OMAprime},~\eqref{SI:eq:Ktrans}, and ~\eqref{SI:eq:Oprime}.
First, combining the Kerr effect with the optomechanical interaction gives
\begin{equation}\label{SI:eq:aprime2}
\begin{split}
& a' = - \frac{i}{\sqrt{3}} a^\dag e^{2i\theta} \sin{\bar{v}} + a (\cos{\bar{v}} - \frac{2i}{\sqrt{3}}\sin{\bar{v}}) + (\frac{1}{3}\cos{\bar{v}} - \frac{i}{\sqrt{3}}\sin{\bar{v}} - \frac{1}{3}) e^{i\theta} r - i\sqrt{2}\chi u  e^{i\theta} \hat{X}(\varphi + \frac{\pi}{2}),\\
& b' = b - \sqrt{2} \chi u  e^{i\varphi} [\hat{x}(\theta)+r].
\end{split}
\end{equation}
Next, after the optical displacement, the final transformation is obtained as 
\begin{equation}\label{SI:eq:Totalaprime}
\begin{split}
& a' = - \frac{i}{\sqrt{3}} a^\dag e^{-i[(\tau_0+\tau)\omega\s{o}-2\theta]} \sin{\bar{v}} 
+ a e^{i(\tau_0+\tau)\omega\s{o}} (\cos{\bar{v}} - \frac{2i}{\sqrt{3}}\sin{\bar{v}}) \\
& \qquad + re^{i \theta} \left[ \left(e^{i(\tau_0+\tau)\omega\s{o}} - \frac{2}{3} \right) \cos{\bar{v}} - \frac{i}{\sqrt{3}} \left(e^{-i(\tau_0+\tau)\omega\s{o}}+ 2 e^{i(\tau_0+\tau)\omega\s{o}} -2 \right)\sin{\bar{v}}  -\frac{1}{3} \right] \\
&\qquad - i\sqrt{2}\chi u  e^{i\theta} \hat{X}(\varphi + \frac{\pi}{2}),\\
& b' = b - \sqrt{2} \chi u  e^{i\varphi} [\hat{x}(\theta)+r].
\end{split}
\end{equation}

In Eq.~\eqref{SI:eq:Totalaprime}, we have considered the most general situation to obtain the transformations. However, these expressions can be simplified by some general considerations.
One instance, presented in the main text, is as follows. The initial phase of the optical mode, $\theta$, is being set by the internal clock of the probe laser which will be further used for the homodyne measurement of the light.
Therefore, we can set $\theta = 0$, that is $\alpha = r \in \mathbb{R}$. Consequently,
\begin{equation}\label{SI:eq:TotalaprimeSim1}
\begin{split}
& a' = - \frac{i}{\sqrt{3}} a^\dag e^{-i(\tau_0+\tau)\omega\s{o}} \sin{\bar{v}} 
+ a e^{i(\tau_0+\tau)\omega\s{o}} (\cos{\bar{v}} - \frac{2i}{\sqrt{3}}\sin{\bar{v}}) \\  
& \qquad + r \left[ \left(e^{i(\tau_0+\tau)\omega\s{o}} - \frac{2}{3} \right) \cos{\bar{v}} - \frac{i}{\sqrt{3}} \left(e^{-i(\tau_0+\tau)\omega\s{o}}+ 2 e^{i(\tau_0+\tau)\omega\s{o}} -2 \right)\sin{\bar{v}}  -\frac{1}{3} \right] \\&\qquad - i\sqrt{2}\chi u  \hat{X}(\varphi + \frac{\pi}{2}),\\
& b' = b - \sqrt{2} \chi u  e^{i\varphi} [\hat{x}+r].
\end{split}
\end{equation}
Another simplification can be obtained by noting that one can arrange the delay for the displacement of the output optical field ($\tau_0$) with high accuracy in such a way that $(\tau_0 + \tau)\omega\s{o} = m 2\pi$ for $m \in \mathbb{N}$. This gives
\begin{equation}\label{SI:eq:TotalaprimeSim4}
\begin{split}
& a' = - \frac{i}{\sqrt{3}} a^\dag \sin{\bar{v}} 
+ a (\cos{\bar{v}} - \frac{2i}{\sqrt{3}}\sin{\bar{v}}) \\  
& \qquad + r \left( \frac{1}{3} \cos{\bar{v}} - \frac{i}{\sqrt{3}} \sin{\bar{v}}  -\frac{1}{3} \right) \\&\qquad - i\sqrt{2}\chi u  \hat{X}(\varphi + \frac{\pi}{2}),\\
& b' = b - \sqrt{2} \chi u  e^{i\varphi} [\hat{x}+r].
\end{split}
\end{equation}
Finally, if we choose the probe pulse duration $\tau$ in such a way that $\sin(\bar{v}) = 0$, we will obtain the simplest transformation equations.
For this, we recall that $\bar{v} = 2\sqrt{3} v = 2\sqrt{3} \chi^2[\omega\s{m}\tau  - \sin(\omega\s{m}\tau )]$.
As $\tau \neq 0 \Rightarrow \bar{v} \neq 0$, we must have
\begin{equation}\label{SI:eq:cond}
\bar{v} = k\pi,  \quad \text{for} \quad k \in \mathbb{N} \qquad \Rightarrow \qquad \omega\s{m}\tau  - \sin(\omega\s{m}\tau ) = \frac{k\pi}{2\sqrt{3} \chi^2}.
\end{equation}
In general, it is always possible to find a pair of parameters $(\tau,r)$, such that $0 \ll \omega\s{m}\tau \ll 2\pi$, and Eq.~\eqref{SI:eq:cond} is satisfied. However, as indicated earlier, the optimal pulse length to achieve $u=2$ is $\omega\s{m}\tau = \pi$. In this case, the condition in Eq.~\eqref{SI:eq:cond} can be satisfied by adjust the amplitude $r$ of the optical coherent state, and thus the parameter $\chi$.
Chosing $\tau$ such that $\bar{v}=k\pi$, we obtain
\begin{equation}\label{SI:eq:TotalaprimeSim5}
\begin{split}
&\left\{\begin{matrix*}[l]
a' = - a  - \frac{2}{3} r - i\sqrt{2}\chi u \hat{X}(\varphi + \frac{\pi}{2}) & \quad k\text{~odd}\\ 
a' = a - i\sqrt{2}\chi u \hat{X}(\varphi + \frac{\pi}{2}) & \quad k\text{~even}
\end{matrix*}\right.
,\\
& \quad b' = b - \sqrt{2} \chi u  e^{i\varphi} [\hat{x}+r],
\end{split}
\end{equation}
where $u$ and $\varphi$ are given by Eq.~\eqref{SI:eq:muparam}.
We see that, for even $k$, the displacement term completely goes away, and we arrive at Eq.~\eqref{eq:aprime}, that is,
\begin{equation}\label{SI:eq:TotalaprimeSim3}
\begin{split}
& a' = a - i\sqrt{2}\chi u \hat{X}(\varphi + \frac{\pi}{2}),\\
& b' = b - \sqrt{2} \chi u  e^{i\varphi} [\hat{x}+r].
\end{split}
\end{equation}

To summarize our parameter choices, consider the case where we choose the optimal pulse length $\tau$ such that $\omega\s{m}\tau=\pi$. This implies $u=2$ and by Eq.~\eqref{SI:eq:cond}, for example for $k=32$, $\chi=\sqrt{\frac{16}{\sqrt{3}}}\approx 3.04$, which is within the required range for Wigner function tomography using our method, without exploiting squeezing. Alternatively, if one does not have precise control over the amplitude $r$ of the input optical coherent state, then for any value of $\chi\sim 1-10$, one can find a pulse length that satisfies the condition in Eq.~\eqref{SI:eq:cond} with $u\neq 0$, for the purpose of Wigner function tomography.
Importantly, for any pair of parameters $(\tau,r)$ satisfying Eq.~\eqref{SI:eq:cond} such that $0 \ll \omega\s{m}\tau \ll 2\pi$, a $s$-parametrized quasiprobability distribution of the mechanics is possible, as described in the main text.

From Eq.~\eqref{SI:eq:TotalaprimeSim3}, after disregarding the mechanical constant displacement, we find the following symplectic matrices,
\begin{equation}\label{SI:eq:SMatrixOMsimplified}
\begin{split}
& S = \begin{pmatrix}
 1 & -\chi u  e^{-i\varphi} & 0 & -\chi u e^{i\varphi} \\ 
 -\chi u e^{i\varphi} & 1 & \chi u e^{i\varphi} & 0 \\
 0 & -\chi u  e^{-i\varphi} & 1 & -\chi u  e^{i\varphi}\\
 \chi u e^{-i\varphi} & 0 & -\chi u e^{-i\varphi} & 1
\end{pmatrix} , \qquad 
S^{-1} = \begin{pmatrix}
 1 & \chi u  e^{-i\varphi} & 0 & \chi u e^{i\varphi} \\ 
 \chi u e^{i\varphi} & 1 & -\chi u e^{i\varphi} & 0 \\
 0 & \chi u  e^{-i\varphi} & 1 & \chi u  e^{i\varphi}\\
 -\chi u e^{-i\varphi} & 0 & \chi u e^{-i\varphi} & 1
\end{pmatrix} ,\\
& \det{S} = 1.
\end{split}
\end{equation}

\section{Proof of Lemma~1}\label{app:ProofLem}

\textbf{Lemma~1.}
\textit{
A product of Weyl-Wigner operators remains a product under all linear transformations of the mode operators, $
\begin{pmatrix}
a'^\dag & b'^\dag & a' & b'
\end{pmatrix}
=
[
\begin{pmatrix}
a^\dag & b^\dag & a & b
\end{pmatrix}
S
+
\begin{pmatrix}
D\s{a}^* & D\s{b}^* & D\s{a} & D\s{a}
\end{pmatrix}
]$,
the argument of which changes according to the inverse of the corresponding symplectic transformation, i.e., $\hat{T}(\alpha)\otimes\hat{T}(\beta)=\hat{T}(\alpha')\otimes\hat{T}(\beta')$ where $\begin{pmatrix}
\alpha'^* & \beta'^* & \alpha' & \beta'
\end{pmatrix}
=
[
\begin{pmatrix}
\alpha^* & \beta^* & \alpha & \beta
\end{pmatrix}
-
\begin{pmatrix}
D\s{a}^* & D\s{b}^* & D\s{a} & D\s{a}
\end{pmatrix}
]
S^{-1}$.
}

\begin{proof}
It is sufficient to prove the theorem for the transformations of Wigner operators. A two-mode Wigner operator is given by~\citep{GlauberBook}
\begin{equation}
\begin{split}
\hat{T}(\alpha)\otimes\hat{T}(\beta) & = \ddmu\xi\zeta e^{\alpha\xi^*-\alpha^*\xi+\beta\zeta^*-\beta^*\zeta} \hat{D}(\xi)\otimes\hat{D}(\zeta) \\
& = \ddmu\xi\zeta \exp\{
\begin{pmatrix}
\alpha^* & \beta^* & \alpha & \beta
\end{pmatrix}
\begin{pmatrix}
-\xi \\
-\zeta \\
\xi^*\\
\zeta^*
\end{pmatrix}
 -
\begin{pmatrix}
a^\dag & b^\dag & a & b
\end{pmatrix}
\begin{pmatrix}
-\xi \\
-\zeta \\
\xi^*\\
\zeta^*
\end{pmatrix}
\}.
\end{split}
\end{equation}
Now, transforming the operators via a linear transformation $\hat{\mathcal{U}}$ and making use of its correspondence with a symplectic group element $S$ (with $|\det S| = 1$), we have
\begin{equation}
\begin{split}
&\hat{\mathcal{U}}\hat{T}(\alpha)\otimes\hat{T}(\beta)\hat{\mathcal{U}}^\dag \\
& = \ddmu\xi\zeta e^{\alpha\xi^*-\alpha^*\xi+\beta\zeta^*-\beta^*\zeta} \hat{\mathcal{U}}\hat{D}(\xi)\otimes\hat{D}(\zeta)\hat{\mathcal{U}}^\dag \\
& = \ddmu\xi\zeta \exp\{
\begin{pmatrix}
\alpha^* & \beta^* & \alpha & \beta
\end{pmatrix}
\begin{pmatrix}
-\xi \\
-\zeta \\
\xi^*\\
\zeta^*
\end{pmatrix}
 -
[
\begin{pmatrix}
a^\dag & b^\dag & a & b
\end{pmatrix}
S
+
\begin{pmatrix}
D\s{a}^* & D\s{b}^* & D\s{a} & D\s{a}
\end{pmatrix}
]
\begin{pmatrix}
-\xi \\
-\zeta \\
\xi^*\\
\zeta^*
\end{pmatrix}
\}\\
& = \ddmu\xi\zeta \exp\{
[
\begin{pmatrix}
\alpha^* & \beta^* & \alpha & \beta
\end{pmatrix}
- \begin{pmatrix}
D\s{a}^* & D\s{b}^* & D\s{a} & D\s{a}
\end{pmatrix}
]
S^{-1}S
\begin{pmatrix}
-\xi \\
-\zeta \\
\xi^*\\
\zeta^*
\end{pmatrix}
 -
\begin{pmatrix}
a^\dag & b^\dag & a & b
\end{pmatrix}
S
\begin{pmatrix}
-\xi \\
-\zeta \\
\xi^*\\
\zeta^*
\end{pmatrix}
\}\\
& = \frac{1}{|\det S|} \ddmu{\xi'}{\zeta'} e^{\alpha'\xi'^*-\alpha'^*\xi'+\beta'\zeta'^*-\beta'^*\zeta'} \hat{D}(\xi')\otimes\hat{D}(\zeta')\\
& = \hat{T}(\alpha')\otimes\hat{T}(\beta'),
\end{split}
\end{equation}
in which
\begin{equation} \label{SI:eq:TransRel}
\begin{pmatrix}
-\xi' \\
-\zeta' \\
\xi'^*\\
\zeta'^*
\end{pmatrix} = 
S
\begin{pmatrix}
-\xi \\
-\zeta \\
\xi^*\\
\zeta^*
\end{pmatrix},\quad
\begin{pmatrix}
\alpha'^* & \beta'^* & \alpha' & \beta'
\end{pmatrix}
=
[
\begin{pmatrix}
\alpha^* & \beta^* & \alpha & \beta
\end{pmatrix}
-
\begin{pmatrix}
D\s{a}^* & D\s{b}^* & D\s{a} & D\s{a}
\end{pmatrix}
]
S^{-1}
\end{equation}
Clearly, this result also applies to single mode linear transformation.
We also bare in mind that $d^2\alpha = d{\rm Im}\alpha \times d{\rm Re}\alpha = \frac{1}{2}d\alpha d\alpha^*=\frac{1}{2}dqdp$.
\end{proof}

\section{Order shift relation}\label{app:OrderShift}
Here we show that the overlap of two quasiprobability distributions with order mismatch $\Delta$ is invariant under exchange of the mismatch between the arguments of either distribution.
We proceed as follows.
\begin{equation}\label{eq:ordershift}
\begin{split}
A & = \dmu\alpha \mathscr{W}_{\hat{\Lambda}}(\alpha;s+\Delta)\mathscr{W}_{\hat{\Upsilon}}(\alpha;t) \\
& = \dmu\alpha {\rm Tr} \hat{\Lambda} \hat{T}(\alpha;s+\Delta) \cdot {\rm Tr} \hat{\Upsilon}\hat{T}(\alpha;t)\\
& = \dmu\alpha \left[\frac{2}{-\Delta}\dmu\xi e^{-\frac{2|\alpha - \xi|^2}{-\Delta}} {\rm Tr} \hat{\Lambda}\hat{T}(\xi;s)\right] \cdot {\rm Tr} \hat{\Upsilon}\hat{T}(\alpha;t)\\
& =  \dmu\xi {\rm Tr} \hat{\Lambda}\hat{T}(\xi;s) \cdot \left[\frac{2}{-\Delta} \dmu\alpha e^{-\frac{2|\alpha - \xi|^2}{-\Delta}} {\rm Tr} \hat{\Upsilon}\hat{T}(\alpha;t)\right]\\
& = \dmu\xi {\rm Tr} \hat{\Lambda}\hat{T}(\xi;s) \cdot {\rm Tr} \hat{\Upsilon}\hat{T}(\xi;t+\Delta).
\end{split}
\end{equation}
In the third and fourth lines we have used the relation~\citep{GlauberBook}
\begin{equation}
\hat{T}(\alpha;s+\Delta) = \frac{2}{-\Delta}\dmu\xi e^{-\frac{2|\alpha - \xi|^2}{-\Delta}}\hat{T}(\xi;s).
\end{equation}

\section{The case of \emph{s}-parameterized tomograms}\label{app:sNonclass}
In the main text we have shown how a single Wigner-function tomogram (i.e.\ $s=0$) of the mechanical state---obtained using our method in the strong interaction regime---can be used to detect the state's P-function nonclassicality. In the case of weaker interactions one generally obtains smoothed $s$-parametrized tomograms of the mechanical state from which a $s$-parametrized quasiprobability distribution can be reconstructed. We will now generalize the criterion of Park \textit{et al}~\citep{Park2017} to $s$-parametrized tomograms and show that even with weak interactions, one can verify the P-function nonclassicality of mechanical states using our method.

Rewriting Eq.~\eqref{eq:Pfunc}, in terms of $s$-parameterized quasiprobability distributions using Eq.~\eqref{eq:Wrep} we have
\begin{equation}\label{eq:sordP}
W_{\hat{\varrho}}(\beta;s) = \dmu\alpha P(\alpha) W_{\alpha}(\beta;s),
\end{equation}
where $W_{\alpha}(\beta;s) = 2 \exp\{ - 2|\beta - \alpha|^2/(1-s) \} /(1-s)$ represents the $s$-parameterized distribution corresponding to the coherent state $\ket{\alpha}$.
We now make the observation that, in the position-momentum coordinates, 
\begin{equation}\label{eq:cohsep}
\begin{split}
W_{\alpha}(q,p;s) = w_\alpha(q,0;s) w_\alpha(p,\frac{\pi}{2};s) \equiv W_{q_0,p_0}(q,p;s),
\end{split}
\end{equation}
where $w_\alpha(q,0;s)= e^{ \frac{-(q - q_0)^2}{1-s}}/\sqrt{\pi(1-s)}$ and $w_\alpha(p,\frac{\pi}{2};s)= e^{ \frac{-(p - p_0)^2}{1-s}}/\sqrt{\pi(1-s)}$ are the two $s$-parameterized tomograms of the coherent state $\ket{\alpha}$ in which $q_0=\sqrt{2}{\rm Re}\alpha$ and $p_0=\sqrt{2}{\rm Im}\alpha$. 
Note that, without loss of generality, we have assumed $\phi=0$. 
Substituting Eq.~\eqref{eq:cohsep} into Eq.~\eqref{eq:sordP}, we obtain
\begin{equation}
\begin{split}
W_{\hat{\varrho}}(q,p;s) & = \int dq'dp' P(q',p') W_{q',p'}(q,p;s) \\
& = \frac{1}{\pi(1-s)}\int dq'dp' P(q',p') e^{ \frac{-(q - q')^2}{1-s}} e^{ \frac{-(q - q')^2}{1-s}}.
\end{split}
\end{equation}
Consequently, the $s$-parameterized tomogram of the state $\hat{\varrho}$, using Eqs.~\eqref{eq:sordP} and~\eqref{eq:cohsep}, is given by
\begin{equation}
\begin{split}
w(q,0;s) & = \frac{1}{\sqrt{\pi(1-s)}}\int dq' \left( \int dp' P(q',p') \right)  e^{ \frac{-(q - q')^2}{1-s}}\\
& = \frac{1}{\sqrt{\pi(1-s)}}\int dq' F(q')  e^{ \frac{-(q - q')^2}{1-s}},
\end{split}
\end{equation}
with $F(q')=\int dp' P(q',p')$.
Importantly, if the state $\hat{\varrho}$ is a classical state, then $P(q',p')\geqslant 0$ is a function with finite support and so is $F(q')$.
Now, suppose that
\begin{equation}\label{SI:eq:sFicTomo}
w\s{f}(p,\frac{\pi}{2};s)  = \frac{1}{\sqrt{\pi(1-s)}}\int dp' G(p')  e^{ \frac{-(p - p')^2}{1-s}},
\end{equation}
where $G(p') \geqslant 0 $ is a positive function with a finite support.
We proceed by defining the fictitious $s$-parameterized distribution $w(x,0;s) \mapsto W\s{f}(q,p;s) = w(x,0;s)w\s{f}(p,\frac{\pi}{2};s)$.
Hence,
\begin{equation}\label{eq:fictsGen}
\begin{split}
W\s{f}(q,p;s) = \frac{1}{\pi(1-s)}\int dq' dp' F(q')G(p') e^{ \frac{-[(q - q')^2+(p - p')^2]}{1-s}}.
\end{split}
\end{equation}
A simple comparison between Eqs.~\eqref{eq:fictsGen} and~\eqref{eq:sordP} shows that $W\s{f}(q,p;s)$ corresponds to a \textit{bona fide} P-classical quantum state, whenever the state $\hat{\varrho}$ is classical.
This is because, in this case, the new function $P\s{f}(q',p')=F(q')G(p')$ is a positive (normalized) function with finite support.
Therefore, it represents a valid P-function. 
Consequently, whenever $W\s{f}(q,p;s)$ fails to represent a \textit{bona fide} fictitious quantum state, we conclude that the initial state $\hat{\varrho}$ was necessarily nonclassical.

Two choices for the fictitious tomogram $w\s{f}(p,\frac{\pi}{2};s)$, as presented in the main text in Eqs.~\eqref{eq:sfic1} and~\eqref{eq:sfic2}, are
\begin{enumerate}[(i)]
\item that of a vacuum state $w\s{f}(p,\frac{\pi}{2};s)=w_0(p,\frac{\pi}{2};s) = e^{ \frac{-p ^2}{1-s}}/\sqrt{\pi(1-s)}$ corresponding to the map $w(x,0;s) \mapsto W\s{f}(q,p;s) = w(x,0;s)w_0(p,\frac{\pi}{2};s)$; and,
\item the tomogram of the state itself, i.e., $w\s{f}(p,\frac{\pi}{2};s)=w(p,0;s)$ corresponding to $(w(x,0;s),w(x,0;s)) \mapsto W\s{f}(q,p;s) = w(x,0;s)w(p,0;s)$.
\end{enumerate}

To check the legitimacy of the fictitious density operators resulting from these demarginalization maps, one may follow the standard arguments provided in Ref.~\citep{Park2017}, such as the Kastler-Loupias-Miracle-Sole test~\citep{Kastler1965,Loupias1966,Nha2008}.
In the general formalism, to perform such tests one may also require evaluation of various Fock basis elements of the fictitious density matrix.
To this end, one should use the trace relation of Eq.~\eqref{eq:trace} and the corresponding $s$-parametrized Fock basis projections given by~\citep{Shahandeh2013}
\begin{equation*}
W_{\ket{n}\bra{m}}(\alpha;s)=(\frac{2}{1-s})^{m+n+1}\frac{1}{\sqrt{n!m!}}  h_{n,m}(\alpha^*,\alpha|\frac{s^2-1}{4}) e^{-\frac{2|\alpha|^2}{1-s}},
\end{equation*}
in which $h_{n,m}(x,y|\epsilon)$ are the incomplete 2D Hermite polynomials defined as~\citep{Dattoli2003}
\begin{equation*}
h_{n,m}(x,y|\epsilon) = \sum_{i=0}^{\min\{m,n\}} \binom{n}{i}\binom{m}{i} i! \epsilon^i x^{n-i} y^{m-i}.
\end{equation*}

\section{The effect of classical noise and losses on nonclassicality certification}
\label{app:Noise}
As discussed in the main text any noise and losses \emph{before} the optomechanical interaction can be accounted for by calibrating the input optical field. We now focus on the readout channel \emph{after} the interaction and show that all noise and losses commute with the interaction and can thus be taken into account as before.
We start by rewriting the Gaussian optical kernel on the r.h.s.\ of Eq.~\eqref{eq:GkernelTomo} as
\begin{equation}
\mathscr{G}\s{o}({\bf x}',{\bf x}_0;{\bf \Sigma}) = \frac{1}{\pi}\exp\{ -\frac{1}{2} ({\bf x'}-{\bf x}_0){\bf \Sigma}^{-1} ({\bf x'}-{\bf x}_0)^{\s{T}}\},
\end{equation}
where ${\bf x'}=(x',p')$, ${\bf x}_0 = (0, X(\varphi+\varphi\s{d}+\frac{\pi}{2}))$, and ${\bf \Sigma}^{-1}={\rm diag}(e^{-2\varepsilon},4\chi^2 u^2 e^{2\varepsilon})$ represents the covariance matrix, so that
\begin{equation}
\begin{split}
\mathscr{W}\s{out;o}(x',2\chi u p';0) & = \frac{1}{\pi} e^{-e^{-2\varepsilon}x'^2} \dmu\beta e^{-4\chi^2 u^2 e^{2\varepsilon} [p' - X(\varphi+\varphi\s{d}+\frac{\pi}{2})]^2} \mathscr{W}\s{m}(\beta;0)\\
& = \dmu\beta \mathscr{G}\s{o}({\bf x}',{\bf x}_0;{\bf \Sigma}) \mathscr{W}\s{m}(\beta;0).
\end{split}
\end{equation}
Since all classical noise and losses acting on the output optical field correspond to Gaussian channels~\citep{Eisert2005}, they can be described as the action of a Gaussian kernel $\mathscr{G}^{\rm noise}({\bf \tilde{x}}, {\bf x}';\boldsymbol{\sigma})$ centred at $(x',p')$, with covariance matrix $\boldsymbol{\sigma}$. Applying the this kernel to the l.h.s.\ of Eq.~\eqref{eq:GkernelTomo} gives
\begin{equation}\label{SI:eq:NoiseModel}
\begin{split}
\mathscr{W}^{\rm noisy}\s{out;o}(\tilde{x},\tilde{p};0) & = \int dx' dp'  \mathscr{G}^{\rm noise}({\bf \tilde{x}}, {\bf x}';\boldsymbol{\sigma}) \mathscr{W}\s{out;o}(x',2\chi u p';0),\\
& = \int dx' dp' \mathscr{G}^{\rm noise}({\bf \tilde{x}}, {\bf x}';\boldsymbol{\sigma}) \Big[ \dmu\beta \mathscr{G}\s{o}({\bf x}',{\bf x}_0;{\bf \Sigma}) \mathscr{W}\s{m}(\beta;0)\Big] .
\end{split}
\end{equation}
It is now easy to see that noise and losses acting on the output optical field commute with the optomechanical interaction and can thus be considered to act on the input field. Formally, we change the order of integrations in Eq.~\eqref{SI:eq:NoiseModel} to obtain
\begin{equation}\label{SI:eq:GausNoiseConv}
\begin{split}
\mathscr{W}^{\rm noisy}\s{out;o}(\tilde{x},\tilde{p};0) & = \dmu\beta \mathscr{W}\s{m}(\beta;0) \Big[ \int dx' dp' \mathscr{G}^{\rm noise}({\bf \tilde{x}}, {\bf x}';\boldsymbol{\sigma}) \mathscr{G}\s{o}({\bf x}',{\bf x}_0;{\bf \Sigma}) \Big]\\
& = \dmu\beta \mathscr{G}^{\rm noise}_{\s{o}}({\bf \tilde{x}}, {\bf x}_0;\boldsymbol{\sigma}+{\bf \Sigma}) \mathscr{W}\s{m}(\beta;0).
\end{split}
\end{equation}
The latter equality follows from the fact that the expression in the square brackets is a convolution of a Gaussian wave packet with a Gaussian kernel which results in a Gaussian wave packet. It is now easy to see that by appropriately rotating the coordinates $(\tilde{x},\tilde{p})$ to diagonalise the covariance matrix $\boldsymbol{\sigma}+{\bf \Sigma}$, one obtains a relation equivalent to Eq.~\eqref{eq:GkernelTomo} and can proceed as before. This shows that classical noise and losses do not distort the picture in our general formalism, assuming that a full characterization of the readout channel is possible.

We would like to emphasize, however, that careful characterization of the readout channel is crucial for \emph{only one} of our nonclassicality criteria, namely the first demarginalization map in Eq.~\eqref{eq:sfic1}. Recall that any Gaussian convolution of the form of Eq.~\eqref{SI:eq:GausNoiseConv} is equivalent to a degradation of a given $s$-parameterized phase-space distribution to another distribution with $s'\leqslant s$. Consequently, since the criterion in Eq.~\eqref{eq:sfic1}, requires that the $s$ parameter of the fictitious tomogram $w\s{f}(p,\frac{\pi}{2};s)$ matches that of the measured tomogram $w\s{f}(x,0;s)$, precise knowledge of the actual experimental value of $s'$ is important. Relaxing this requirement somewhat, it is also possible to use a fictitious tomogram with $s'\leqslant s$ (i.e. an overestimation of the actual noise), because, using Eq.~\eqref{SI:eq:sFicTomo}, one possible fictitious tomogram $w\s{f}(p,\frac{\pi}{2};s')$ is given by
\begin{equation}
w\s{f}(p,\frac{\pi}{2};s')  = \frac{1}{\sqrt{\pi(1-s')}}\int dp' \mathscr{G}^{\rm noise}(p', 0;\sigma)  e^{ \frac{-(p - p')^2}{1-s'}},
\end{equation}
where $\sigma = \frac{s-s'}{(1-s')(1-s)}$ and $\mathscr{G}^{\rm noise}(p', 0;\sigma)$ is a Gaussian with finite support provided that $s'\leqslant s$. Consequently, in the presence of noise and loss, one should be careful in using the nonclassicality criterion in Eq.~\eqref{eq:sfic1}.

Importantly, such a caution is not needed when using the second demarginalization map, Eq.~\eqref{eq:sfic2}. In this case, the experimentally measured tomogram is used for both quadratures and the ordering parameters are thus always matched, regardless of the actual experimental value of $s$. As a result, our second nonclassicality criteria is immune to void nonclassicality detections due to noise and losses.


\end{widetext}

\bibliography{OptomechanicsTomography}
\bibliographystyle{apsrev4-1new}


\end{document}